\documentclass[11pt]{article}
\usepackage[utf8]{inputenc}
\usepackage[margin=1in]{geometry}
\usepackage{amsmath,amssymb}
\usepackage{graphicx}
\usepackage{booktabs}
\usepackage{array}
\usepackage{float}
\usepackage{caption}
\usepackage[expansion=false]{microtype}
\usepackage[dvipsnames]{xcolor}
\usepackage{xspace}
\usepackage[colorlinks=true,linkcolor=NavyBlue,citecolor=NavyBlue,urlcolor=NavyBlue]{hyperref}

\graphicspath{{figures/}}
\newcommand{\fig}[3]{\begin{figure}[htbp]\centering\includegraphics[width=#1\linewidth]{#2}\caption{#3}\label{fig:#2}\end{figure}}
\newcommand{\Fig}[1]{Figure~\ref{fig:#1}}
\newcommand{\degC}{\ensuremath{^{\circ}}C\xspace}

\title{Forecasting Daily Household Electricity Demand in a Hot-Arid City:\\
A Comparative Benchmark of Local, Global, Deep and Foundation Models on ESHED-38}
\author{Bisheshwor Neupane\\
\texttt{neupan10@rowan.edu}}
\date{}

\begin{document}
\maketitle

\begin{abstract}
This study benchmarks 24 forecasting configurations on ESHED-38, a public record of daily smart-meter consumption for 38 homes in Erbil, Iraq, collected between May 2020 and June 2025 together with station weather and a household survey. The configurations span five base families, plus two cross-family ensembles: naive baselines, per-house statistical models, global gradient-boosted trees, global deep networks, and pre-trained time-series foundation models applied zero-shot. Every configuration is evaluated on the same 54{,}720 forecasts, produced from 48 weekly origins across a held-out period of 359 days at lead times of up to 30 days, and scored for point accuracy (scaled absolute error), distributional quality (continuous ranked probability score) and calibration, with Friedman, Nemenyi and Holm-corrected Wilcoxon tests across houses. The best configurations cut the geometric-mean scaled error of a seasonal-naive forecast by 20 to 25 percent, and the leading models beat that baseline in every one of the 38 houses. Chronos-2, used without any training on these data, is the most accurate model one day ahead and is significantly better there than all 23 alternatives; at one week it is statistically tied with TimesFM-2.5, Chronos-Bolt and a cross-family ensemble fixed before testing, and the two ensembles lead at 30 days. Deep networks trained from scratch do not outperform exponential smoothing on average, and the household survey adds negligible skill. Metered demand peaks in winter and falls as temperature rises. Supplying the recorded weather for the forecast days did not improve any model that was run both with and without it. A quality screen helps explain why: on 68 percent of dates at least one temperature value recurs exactly on two or more other dates, and these dates form three contiguous annual blocks, the last covering almost all of the test period. This pattern suggests that much of the weather record was filled by reusing one annual sequence. The value of weather information for these homes is therefore treated as unresolved, and the protocol is released as a baseline for future work.
\end{abstract}


\section{Introduction}

A forecast of how much electricity one home will use tomorrow, or next week, is useful to more people than might be expected. Utilities need it to plan supply and to target demand-response offers; network operators need it to judge how close local transformers run to their limits; and households themselves can use it to anticipate bills. Yet the single home is the hardest unit to forecast. A city's demand averages out thousands of independent decisions, so its daily load follows the weather and the calendar closely. One meter, on the other hand, registers every trip away, every visiting relative and every heater left on overnight.

The research literature on household forecasting draws mainly on data from Europe, North America and East Asia. Homes in the hot-arid Middle East differ from those settings in almost every relevant respect: building materials, the appliances that people own, the way they heat and cool, and, in several countries, the reliability of the grid itself. The Kurdistan Region of Iraq is a clear case. Summer days regularly pass 45\,\degC, winter nights are cold enough to need space heating, and many families draw part of their power from private neighbourhood generators. Whether modern forecasting methods work well under these conditions is largely unknown, mostly because public data have not been available.

The ESHED-38 dataset, released by Jaf and Hamarash in 2026~\cite{eshed38}, fills part of that gap. It records five years of daily smart-meter readings for 38 homes in Erbil, aligned with weather measurements and a survey of building and appliance characteristics. It is used here to address three practical questions. Which family of forecasting methods works best for daily single-home demand in this setting? How much better than a simple rule are the best methods? And can the uncertainty they report be trusted?

Two developments make these questions timely. One is the ongoing comparison between local models, fitted separately to each series, and global models trained once across many series~\cite{montero2021}. The other is the arrival of large pre-trained time-series foundation models, such as Chronos~\cite{chronos} and TimesFM~\cite{timesfm}, which are meant to forecast series they have never seen. Household meters in Iraq are almost certainly absent from the data these models were trained on, which makes ESHED-38 a useful out-of-distribution test.

This paper contributes the following:
\begin{enumerate}
\item To the author's knowledge, the first forecasting benchmark on ESHED-38: 24 configurations from five base model families plus two cross-family ensembles, compared on identical forecasts under a leakage-free rolling-origin protocol with 48 origins and three horizons.
\item An evaluation that covers point accuracy, the quality of the full predictive distribution and calibration, with non-parametric significance tests that treat each house as a paired observation.
\item Practical findings: zero-shot foundation models are among the most accurate options at a negligible computing cost, networks trained from scratch are not, and survey attributes add only negligible signal for this task.
\item A data-quality analysis showing that most of the weather record behaves like a repeated annual sequence rather than a measured daily series, with direct consequences for any study of weather effects on these data.
\end{enumerate}

Section~\ref{sec:related} reviews related work. Section~\ref{sec:data} describes the data, the quality audit and the exploratory analysis. Section~\ref{sec:method} defines the forecasting task and the models, and Section~\ref{sec:eval} the evaluation protocol. Results follow in Section~\ref{sec:results}, discussion and limitations in Section~\ref{sec:discussion}, and conclusions in Section~\ref{sec:conclusion}.


\section{Related Work}\label{sec:related}

\subsection{Household load forecasting}
Load forecasting at the level of utilities and regions has a long history. At those scales demand is smooth and tracks temperature and the calendar closely. Hong and Fan~\cite{hong2016} survey the probabilistic branch of this work and argue that a single point forecast says too little when decisions depend on risk. The spread of smart meters turned attention to individual homes. Kong et al.~\cite{kong2019}, for example, show that recurrent networks can learn some household-level regularities, while also making clear how much the task hardens once consumption is no longer averaged over many customers. Most such studies use hourly or half-hourly data from temperate climates; daily data from hot-arid regions, with their different heating, cooling and supply conditions, have received far less attention.

\subsection{Local and global models}
Exponential smoothing~\cite{hyndman2002}, automatic ARIMA~\cite{hyndman2008} and the Theta method~\cite{theta2000} are fitted one series at a time. Global models instead learn from a pool of series at once. Montero-Manso and Hyndman~\cite{montero2021} show that a global model can match local ones even when the series appear unrelated, provided it has enough capacity. The M4~\cite{m4} and M5~\cite{m5} competitions supported this in practice: globally trained gradient-boosted trees such as LightGBM~\cite{lightgbm} did well in M5, and combinations of different methods ranked near the top in both competitions.

\subsection{Deep and foundation models}
Forecasting-specific deep architectures include DeepAR~\cite{deepar}, an autoregressive recurrent network with a parametric output distribution; the Temporal Fusion Transformer~\cite{tft}; N-HiTS~\cite{nhits}, which forecasts at several time resolutions; and PatchTST~\cite{patchtst}, which applies a transformer to short patches of a series. They usually need large training sets, and their advantage over simpler methods on small, noisy collections remains unclear. Foundation models take a different route. A single large network is pre-trained on a very large and varied corpus of time series and then applied to new series without further training. Chronos~\cite{chronos} quantises scaled values into tokens and trains a language-model architecture on them, while TimesFM~\cite{timesfm} uses a decoder-only transformer over patches. Independent evaluations of such models on household energy data, especially from regions missing from their training corpora, are still rare.

\subsection{Evaluating forecasts}
How forecasts are scored shapes what a study concludes. Hyndman and Koehler~\cite{hyndman2006} introduced the mean absolute scaled error so that accuracy can be compared across series of different size. For probabilistic forecasts, strictly proper scoring rules such as the continuous ranked probability score reward honest statements of uncertainty~\cite{gneiting2007a}, and Gneiting et al.~\cite{gneiting2007b} describe the aim as maximising sharpness subject to calibration. When many methods are compared over many series, Dem\v{s}ar~\cite{demsar2006} recommends rank-based tests rather than averages alone. Tashman~\cite{tashman2000} and Bergmeir and Ben\'itez~\cite{bergmeir2012} stress that out-of-sample evaluation should always forecast forward from a fixed origin. All of these recommendations are followed here.


\section{Data}\label{sec:data}

\subsection{The ESHED-38 dataset}
ESHED-38 (Erbil Smart Home Energy Dataset, version~2) was published on Mendeley Data in May 2026 under a CC~BY~4.0 licence~\cite{eshed38}. It covers 38 houses in Erbil from 14 May 2020 to 17 June 2025, which gives 1{,}861 daily records per house and 70{,}718 records in all. Each record combines three kinds of information:
\begin{itemize}
\item \textbf{Energy:} daily consumption in kWh, from smart meters installed by the Ministry of Electricity.
\item \textbf{Weather:} daily mean, minimum and maximum of air temperature, relative humidity and barometric pressure; wind speed, gusts and direction; and solar irradiance and daily solar energy. According to the data description these come from sensors close to the houses, and on any given date the same values apply to every house.
\item \textbf{Household attributes:} occupancy, floor area, number of rooms, floors and windows, glazing, orientation, shading, floor and wall type, and ownership of 15 appliances, gathered by a trained field team in a survey in December 2025.
\end{itemize}
Because the survey took place after the metering period, household attributes are treated as fixed for each house, with the caveat that they may not describe every home correctly in earlier years.

\subsection{Data quality audit}
Table~\ref{tab:audit} summarises the checks run before any modelling.

\begin{table}[htbp]
\centering\small
\caption{Data quality audit of ESHED-38.}\label{tab:audit}
\begin{tabular}{p{0.47\linewidth}p{0.45\linewidth}}
\toprule
Check & Result\\
\midrule
Records per house & 1{,}861 for each of 38 houses\\
Missing dates, duplicate rows, missing cells & none\\
Negative consumption & none\\
House-days with exactly zero consumption & 90 (53 in runs of three days or more)\\
House-days below 1\,kWh & 424, in 24 of the 38 houses\\
Weather identical across houses on each date & yes\\
Survey attributes constant within each house & yes\\
Daily solar energy versus irradiance & ratio 10.00 (unit error)\\
Irradiance above 1.5\,kW/m$^2$ & 5 dates (values near 90\,kW/m$^2$)\\
Survey columns constant across all houses & 7\\
Survey attributes held by all but one house & oven, boiler\\
\bottomrule
\end{tabular}
\end{table}

The audit found a complete and regular panel, but four issues needed attention. The daily solar energy column is exactly ten times larger than the irradiance column implies, so it was rescaled. Five dates carry irradiance values close to 90\,kW/m$^2$, roughly sixty times the physical maximum at ground level; they were treated as logger errors and replaced by interpolation in time. Seven survey columns (shading, floor type, wall type, washing machine, evaporative cooler, a second refrigerator column and television) take the same value in every house and were dropped. Finally, 424 house-days fall below 1\,kWh, spread across 24 houses. Only 90 of these are exactly zero, so a check for zeros alone understates how often the meter records almost nothing. The house-by-day map in \Fig{03_house_day_heatmap} makes these periods visible as dark blocks. One house shows about two months of near-zero use in mid-2021, another a long low spell in early 2023, and two houses (H05 from September 2024 and H30 from March 2025) have low periods inside the test period. They are most likely supply interruptions or times when the home was empty. They were kept in the data and flagged for the sensitivity analysis in Section~\ref{sec:sens}.

\fig{1.0}{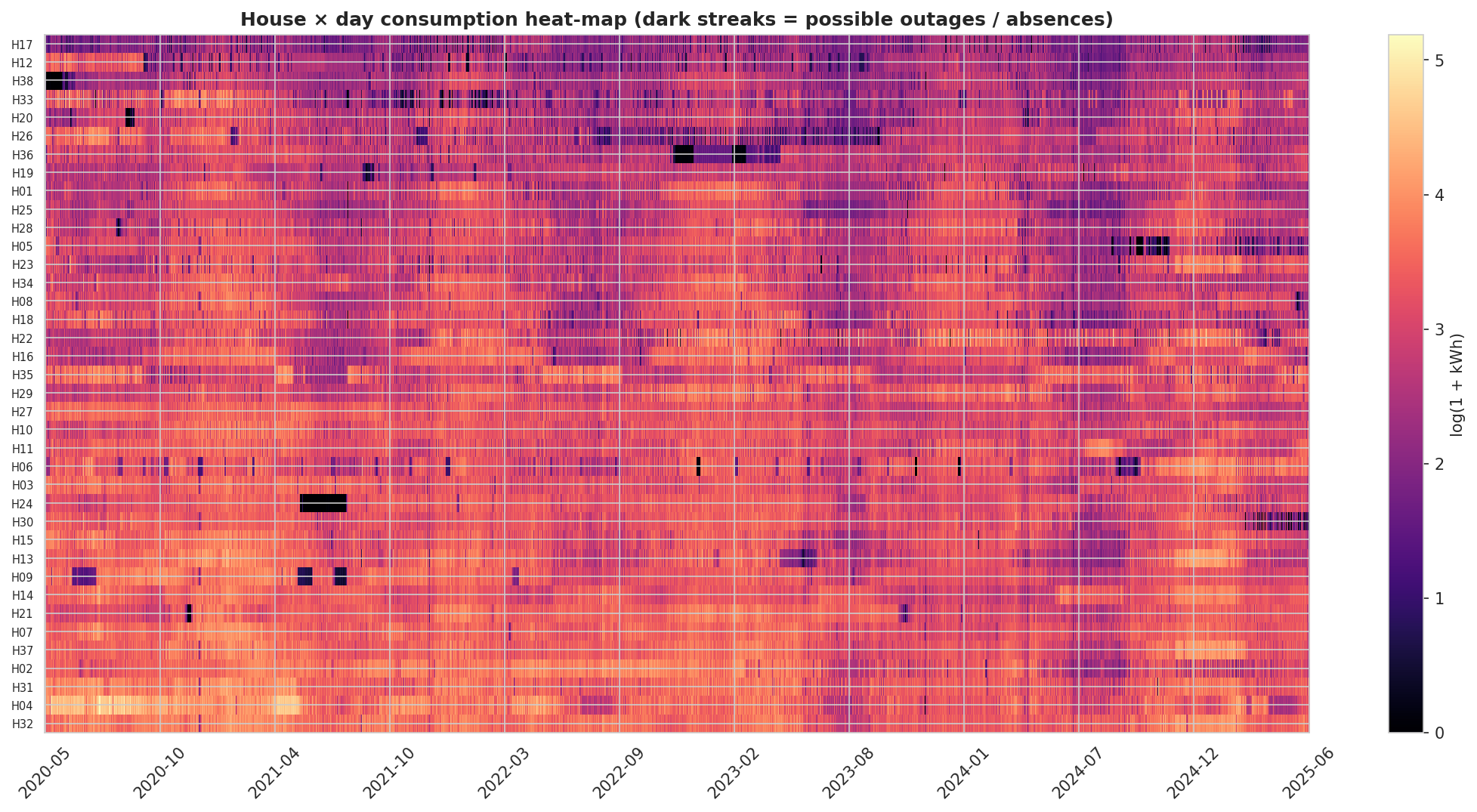}{Daily consumption of every house (rows, ordered by median use) over the full record, on a $\log(1+\text{kWh})$ colour scale. Horizontal dark streaks mark stretches of near-zero use, interpreted as outages or absences. The lighter band in the first year for several houses reflects the higher consumption of 2020--21.}

\subsection{Weather quality}\label{sec:wxqc}
The weather series raised a more serious question. \Fig{08_weather_series} plots each variable over the five years. Several features are hard to reconcile with a measured record: temperature drops repeatedly to the same few values, pressure moves smoothly for months at a time in some summers but fluctuates strongly in others, and the patterns in some years look strikingly alike.

\fig{1.0}{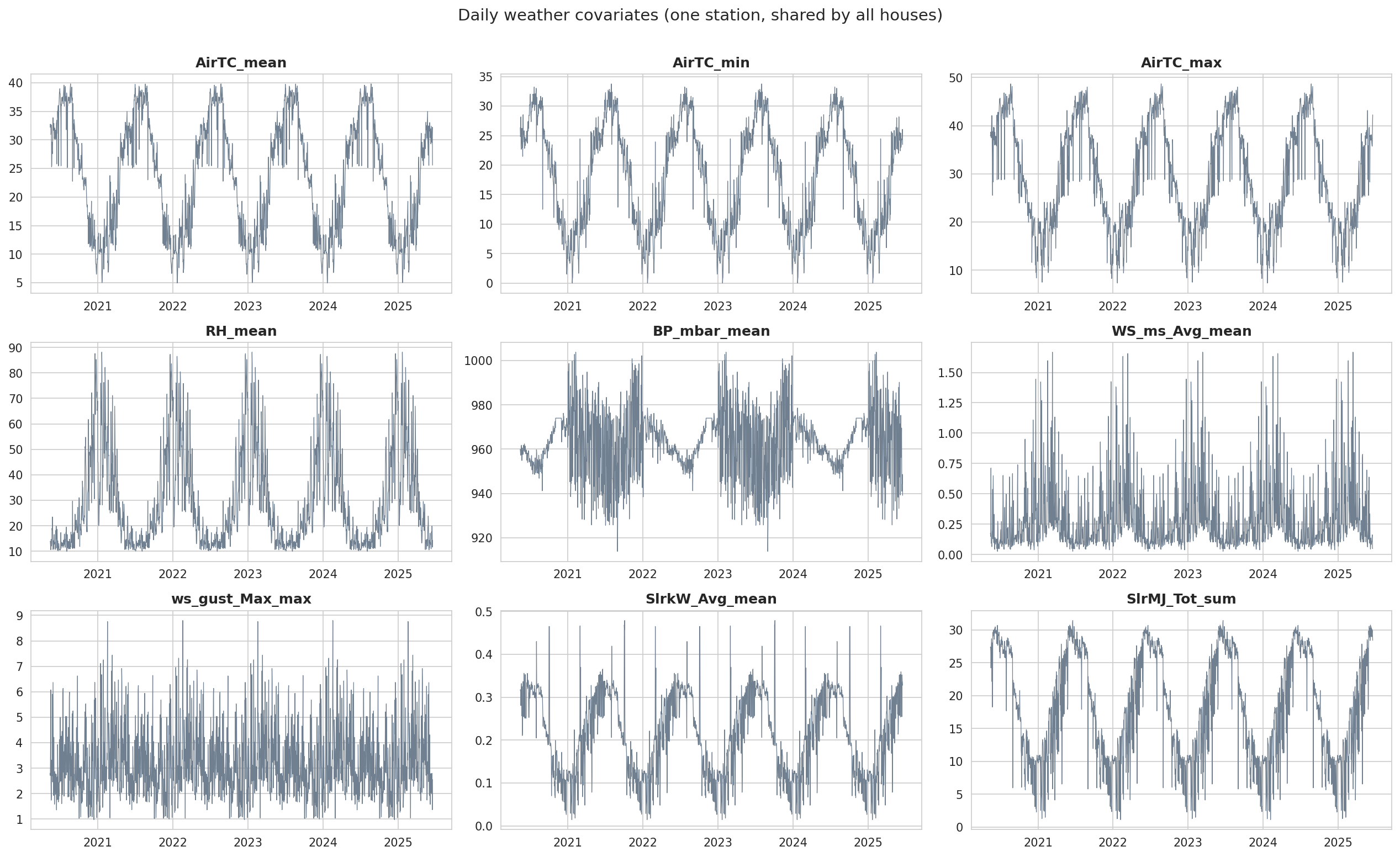}{Daily station weather after the solar corrections: air temperature (mean, minimum, maximum), relative humidity, pressure, mean and gust wind speed, irradiance and daily solar energy. All variables show the expected annual cycle, but note the identical downward spikes in temperature and the contrast between smooth and noisy years in pressure.}

To examine this systematically, four automatic checks were applied to each variable. A date was flagged if its value, rounded to four decimal places, appears on at least three different dates; if it equals the previous or next day's value; if it lies inside a run of three or more days with zero second difference (an exactly straight line); or if the seven-day median of absolute day-to-day changes is below 15 percent of its overall median. The daily means in ESHED-38 are averages over many sub-daily readings and are stored with eight or more decimal places (for example, 32.72614583\,\degC). At that precision two independently measured days should essentially never share a mean value, let alone three. Daily minima and maxima are stored with two decimals, so a few chance repeats are possible for them, but not on the scale observed below.

The result is striking. Between 64 and 74 percent of dates are flagged for each variable (Table~\ref{tab:wxqc}), and almost all of these flags come from the repeated-value check. Taking the three temperature variables together, 1{,}266 of 1{,}861 dates (68 percent) are flagged. \Fig{08b_weather_quality_control} shows where the flags fall. They are not scattered: they form three contiguous blocks, from mid-May 2020 to June 2021, from May 2022 to June 2023, and from May 2024 to the end of the record. Measured month by month on the mean-temperature panel, essentially every date inside these blocks is flagged, against 3 to 5 percent of dates in the two years between them. The repeated-value check requires a value to occur on at least three dates, and there are exactly three blocks. The simplest explanation is therefore that the three blocks are copies of one annual weather sequence, used perhaps to fill gaps in the station data. Whether the years in between are genuine measurements, or a second sequence used only twice (which this check could not detect), cannot be determined from the data. It would not then be an independent daily measurement. This could not be confirmed with the data owners before writing, so it is presented as a strong indication rather than a settled fact. It matters a great deal for this study, because the third block covers almost the entire test period.

\begin{table}[htbp]
\centering\small
\caption{Weather quality screen: number of dates (out of 1{,}861) flagged by each check.}\label{tab:wxqc}
\begin{tabular}{lrrrrr}
\toprule
Variable & Repeated value & Frozen & Interpolated & Over-smooth & Any flag (share)\\
\midrule
Air temperature, mean & 1{,}196 & 9 & 0 & 6 & 1{,}196 (0.64)\\
Air temperature, min & 1{,}249 & 13 & 0 & 12 & 1{,}249 (0.67)\\
Air temperature, max & 1{,}224 & 19 & 0 & 0 & 1{,}224 (0.66)\\
Relative humidity & 1{,}198 & 9 & 0 & 43 & 1{,}214 (0.65)\\
Pressure & 1{,}228 & 96 & 90 & 144 & 1{,}230 (0.66)\\
Wind speed & 1{,}321 & 9 & 0 & 42 & 1{,}329 (0.71)\\
Irradiance & 1{,}368 & 23 & 0 & 35 & 1{,}372 (0.74)\\
Daily solar energy & 1{,}201 & 9 & 0 & 7 & 1{,}201 (0.65)\\
\bottomrule
\end{tabular}
\end{table}

\fig{1.0}{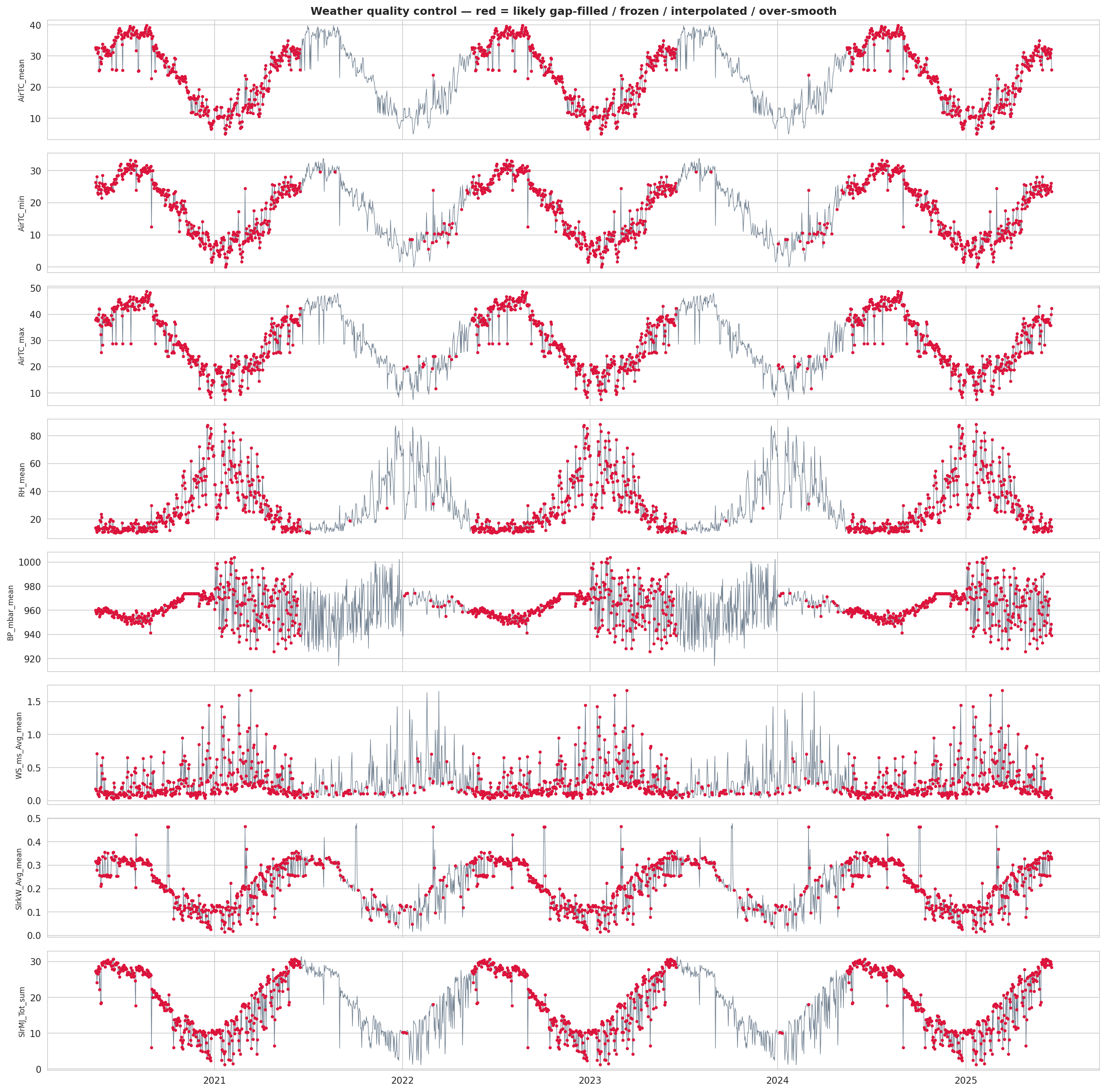}{Weather quality screen. Red points mark dates that fail at least one check (repeated exact value, frozen value, straight-line interpolation or unusually smooth variation). Flagged dates cluster in three annual blocks (mid-May 2020--June 2021, May 2022--June 2023 and May 2024--June 2025), while the years in between pass almost entirely.}

\subsection{Exploratory analysis}\label{sec:eda}

\paragraph{Distribution of daily consumption.} \Fig{01_target_distribution} shows that daily use is strongly right-skewed (skewness 1.16, excess kurtosis 3.40). Most house-days fall between 5 and 50\,kWh, but the tail reaches 178\,kWh. The Q--Q plot departs from normality at both ends, and the log-transformed distribution has two modes, near $e^{2.5}\approx 11$ and $e^{3.3}\approx 26$\,kWh, which reflect two broad groups of low- and high-consuming house-days. Skewness has a practical consequence for forecasting: the mean of the predictive distribution lies above its median, so a model trained to predict the mean will systematically overshoot the median. This point is taken up again in Section~\ref{sec:gbm}.

\fig{1.0}{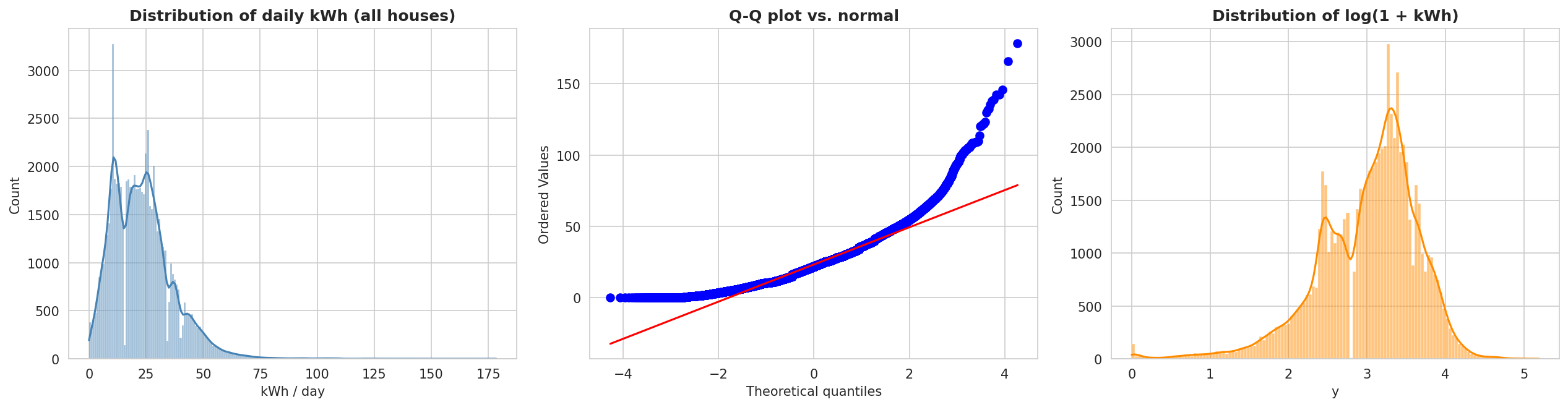}{Distribution of daily consumption over all 70{,}718 house-days: histogram with kernel density (left), normal Q--Q plot (centre) and distribution of $\log(1+\text{kWh})$ (right).}

\paragraph{Differences between houses.} The houses differ greatly in scale (\Fig{02_per_house_boxplot}). Median daily use ranges from about 7\,kWh in H17 to about 36\,kWh in H32, a fivefold spread. Some houses also show long upper tails: H04 exceeds 100\,kWh on many days and reaches 178\,kWh, and H22 reaches 142\,kWh. A single global model must therefore serve series of very different level and volatility. For this reason each series is standardised before the global models are trained, and scale-free metrics are used for evaluation.

\fig{1.0}{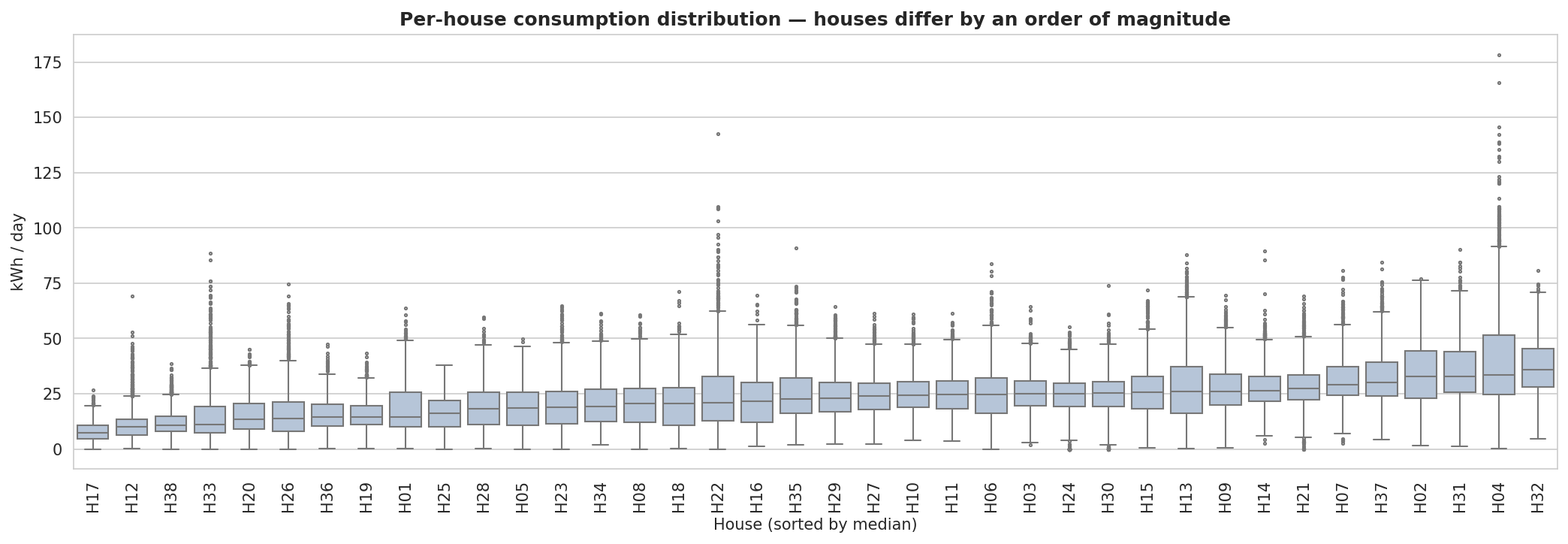}{Daily consumption per house, ordered by median. Boxes show the interquartile range and whiskers 1.5 times that range; points are individual outlying days.}

\paragraph{Demand over time.} \Fig{04_aggregate_series_with_temperature} sets the cross-house mean against air temperature. The two series move in opposite directions. Consumption rises every winter, when temperature falls to around 10\,\degC, and drops every summer, when temperature approaches 38\,\degC. The highest levels occur in the winter of 2020--21, when the 7-day mean exceeds 35\,kWh, and the lowest in mid-2024, around 12--14\,kWh. The spread between the 10th and 90th percentile houses is wide throughout, from below 10\,kWh to 40--60\,kWh.

\fig{1.0}{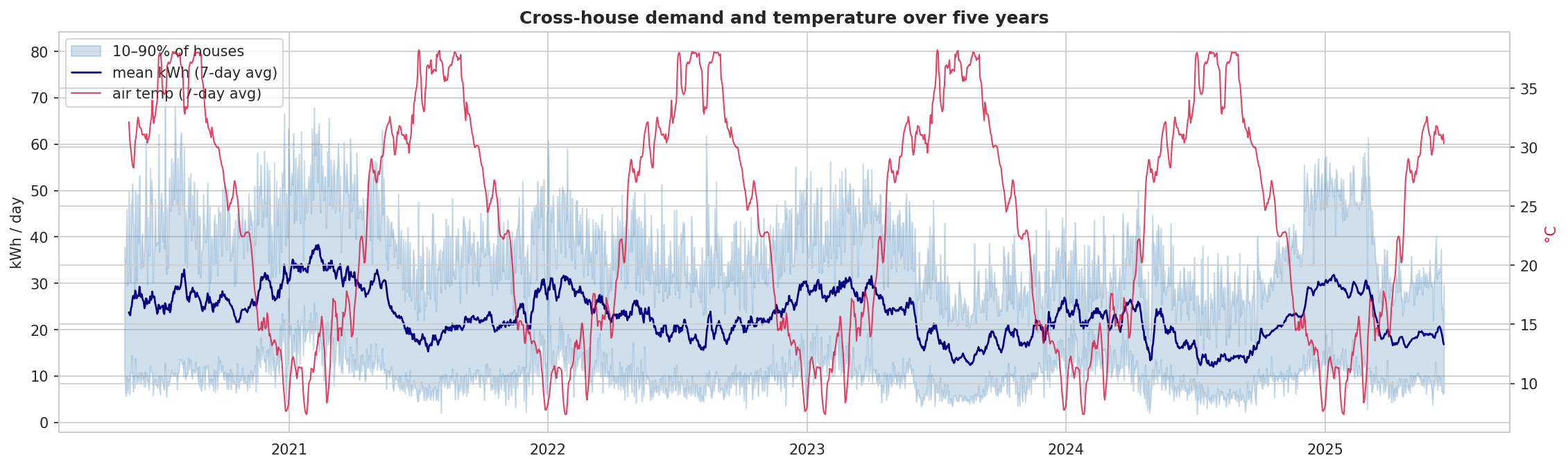}{Seven-day mean consumption across houses (dark line) with the band between the 10th and 90th percentile houses, and the seven-day mean air temperature (red, right axis).}

\paragraph{Response to temperature.} The usual expectation for buildings is a U-shaped response: consumption rises when it is cold, because of heating, and again when it is hot, because of cooling. \Fig{05_energy_signature} shows something different. The smoothed cross-house mean falls steadily from about 29\,kWh per day at 5\,\degC to about 17\,kWh at 40\,\degC, with no sign of a cooling branch. The per-house panel confirms that this is not an artefact of averaging. Almost every house uses 20 to 50 percent more than its own average below 15\,\degC and less than average above 30\,\degC; H35 is one of very few exceptions. Two explanations are plausible. Much summer cooling may run on private generators that the Ministry meter does not record, or on evaporative coolers that draw little power. Winter heating, by contrast, appears to rely on grid-powered electric heaters. The dataset cannot separate these explanations, but either way the metered load is heating-driven. The empty vertical strips in the scatter plot (no days near 15--16, 20--21 or 34--35\,\degC) are consistent with the repeated weather sequences described in Section~\ref{sec:wxqc}.

\fig{1.0}{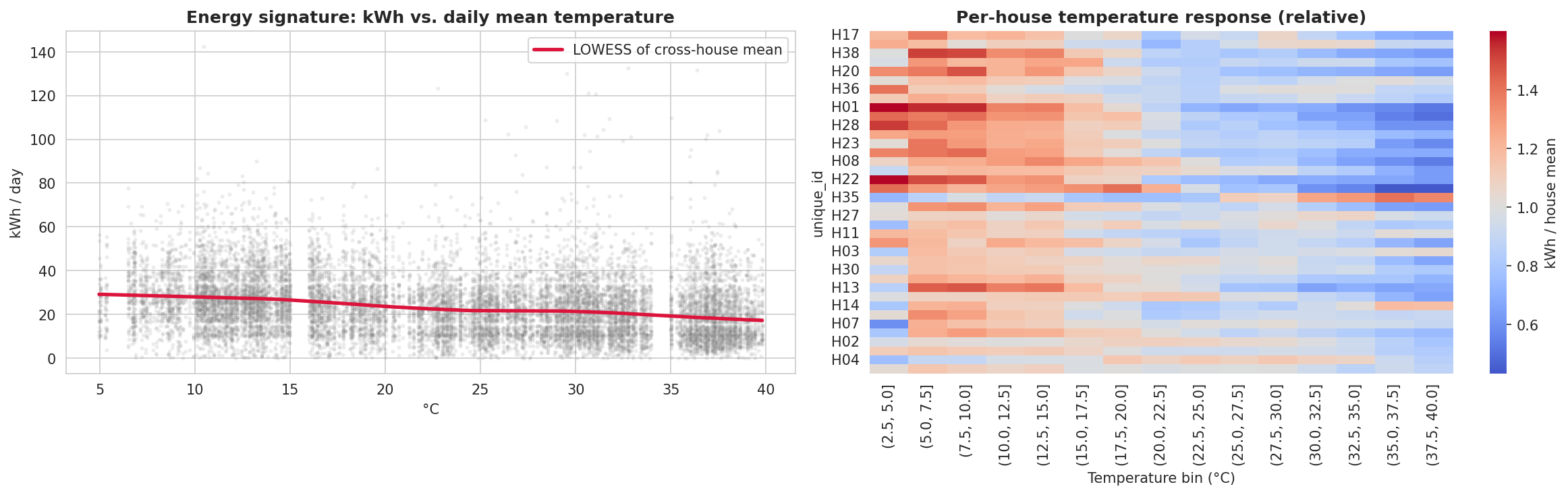}{Left: daily consumption against daily mean temperature for a random sample of house-days, with a LOWESS curve of the cross-house mean. Right: each house's mean consumption in 2.5\,\degC temperature bins, divided by its overall mean (red above, blue below its average).}

\paragraph{Calendar effects.} \Fig{06_calendar_profiles} summarises calendar patterns relative to each house's own mean. The weekly cycle is weak. Friday, the main day of the weekend, is about 6 percent above average and Thursday about 1 percent above, and the other five days lie within 3 percent of the mean. The annual cycle is strong: January (1.27) and February (1.25) are the highest months and July (0.78) the lowest, with a smooth transition through spring and autumn. Among complete years, median house-level consumption falls slightly from about 24\,kWh per day in 2021 and 2022 to about 21\,kWh in 2024. The test period is therefore somewhat calmer than the training years, which affects how scaled errors should be read (Section~\ref{sec:overall}).

\fig{1.0}{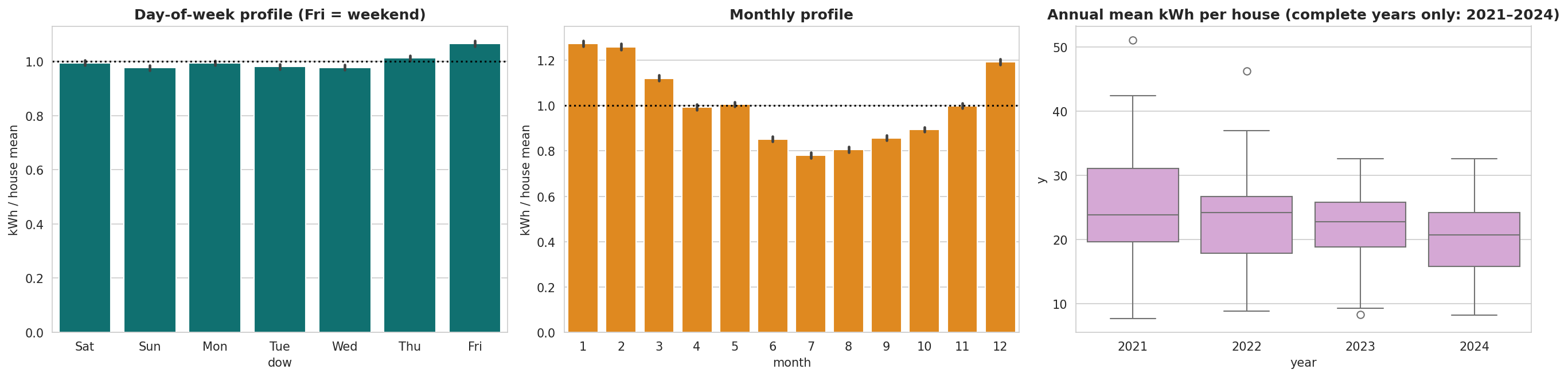}{Day-of-week profile (left) and monthly profile (centre), each relative to the house mean with 95 percent confidence intervals, and the distribution of annual mean consumption per house for the complete years 2021--2024 (right).}

\paragraph{Persistence.} The autocorrelation functions in \Fig{07_acf_pacf} explain why the recent level of consumption is such a strong predictor. For H29 and H32 the autocorrelation stays near 0.5 for the first two weeks and is still about 0.25 after 60 days. Day-to-day use is noisy, but the underlying level drifts slowly. H17, the lowest consumer, is different: its autocorrelation stays between 0.1 and 0.3, so even yesterday says little about today, and any model will struggle with it. Peaks at lags 7, 14 and 21 are present but small, which matches the weak weekly profile. In the partial autocorrelations, the first three lags carry most of the direct dependence.

\fig{1.0}{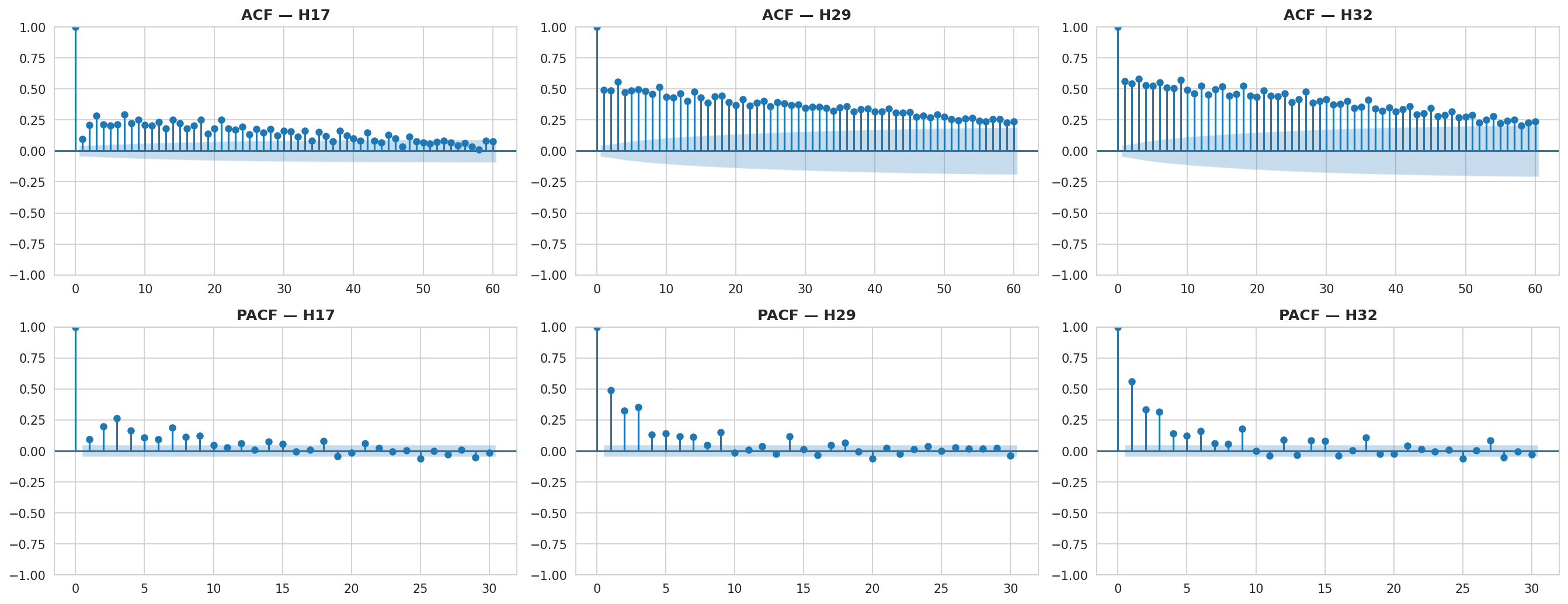}{Autocorrelation (top, 60 lags) and partial autocorrelation (bottom, 30 lags) for a low-consumption (H17), a median (H29) and a high-consumption house (H32). The shaded band is the 95 percent confidence region under no correlation.}

\paragraph{Weather correlations.} \Fig{09_weather_correlation} gives Pearson correlations between the cross-house mean and the daily weather variables, after the solar corrections. Consumption correlates negatively with temperature ($r=-0.53$ for mean and minimum temperature and $-0.52$ for maximum temperature) and with solar energy ($r=-0.41$), and positively with humidity ($r=0.46$), which in Erbil is high in winter. The weather variables are themselves highly collinear. The three temperature measures correlate at 0.93 to 0.97, temperature and humidity at about $-0.8$, and irradiance and daily solar energy at 0.90. Given the findings of Section~\ref{sec:wxqc}, these correlations mainly describe the shared annual cycle rather than day-to-day weather effects.

\fig{0.62}{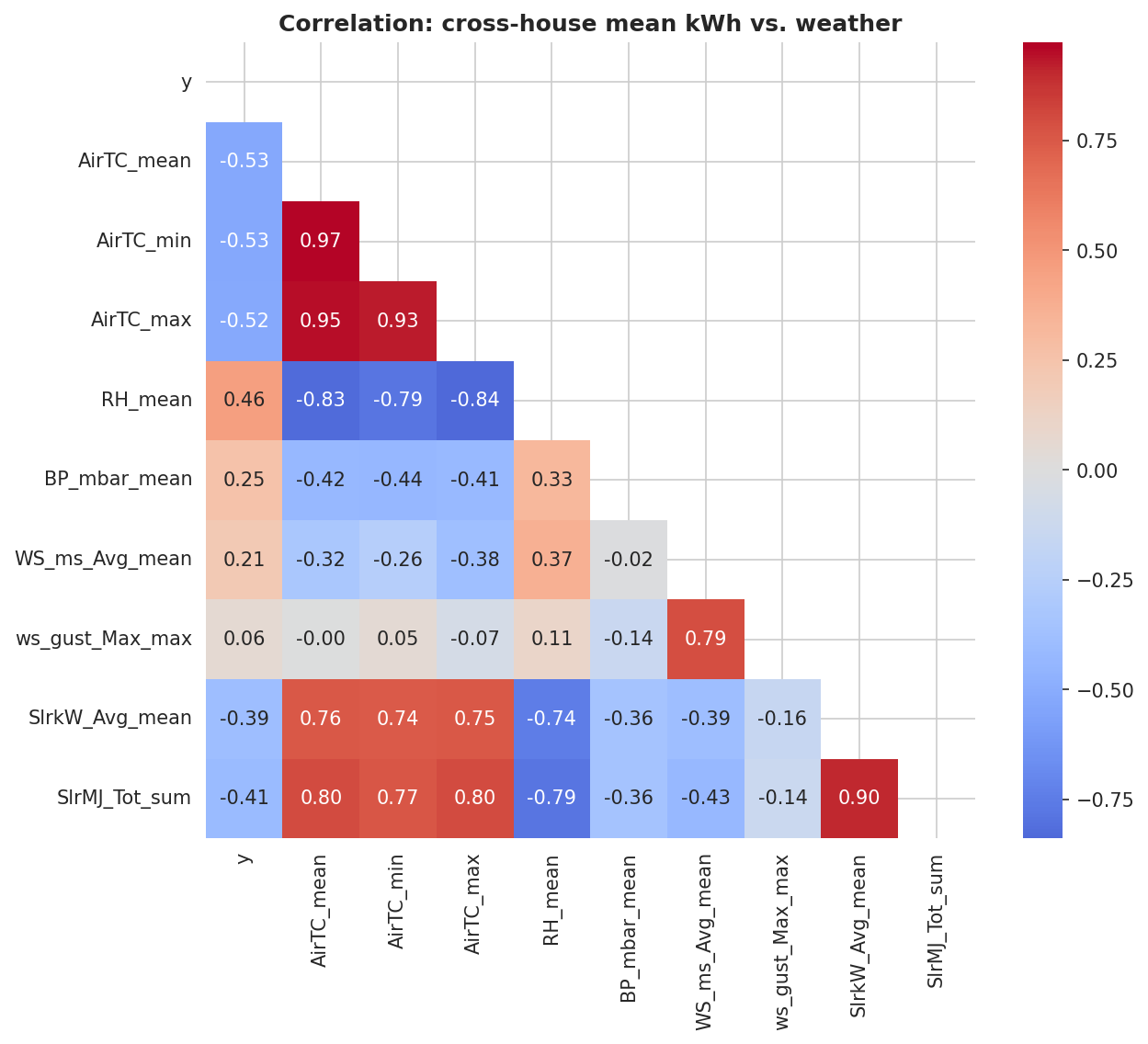}{Pearson correlation between the cross-house mean daily consumption ($y$) and the weather variables.}

\paragraph{Household survey.} \Fig{10_survey_attributes} shows that the survey offers limited contrast between homes. An oven and a water boiler are present in 37 of the 38 houses, so any comparison for them rests on a single home. A refrigerator (82 percent) and fans (79 percent) are nearly universal. Electric heaters (50 percent) and air conditioners (45 percent) split the sample roughly in half, and only three houses (8 percent) use a fuel-fired \emph{shofaz} heater. The right-hand panel gives the raw difference in mean consumption between houses with and without each item. These differences are descriptive, not causal: houses with blackout curtains use about 5\,kWh more per day, and the three \emph{shofaz} houses about 7\,kWh more, but such contrasts are confounded with house size and occupancy. \Fig{11_numeric_attributes_vs_kwh} shows the numeric attributes. The number of rooms is most strongly related to mean consumption ($r=0.51$), followed by occupancy ($r=0.37$) and number of windows ($r=0.35$). Floor area takes only two values, 100 and 200\,m$^2$, with three houses at 200\,m$^2$, which suggests it was recorded as a coarse category.

\fig{1.0}{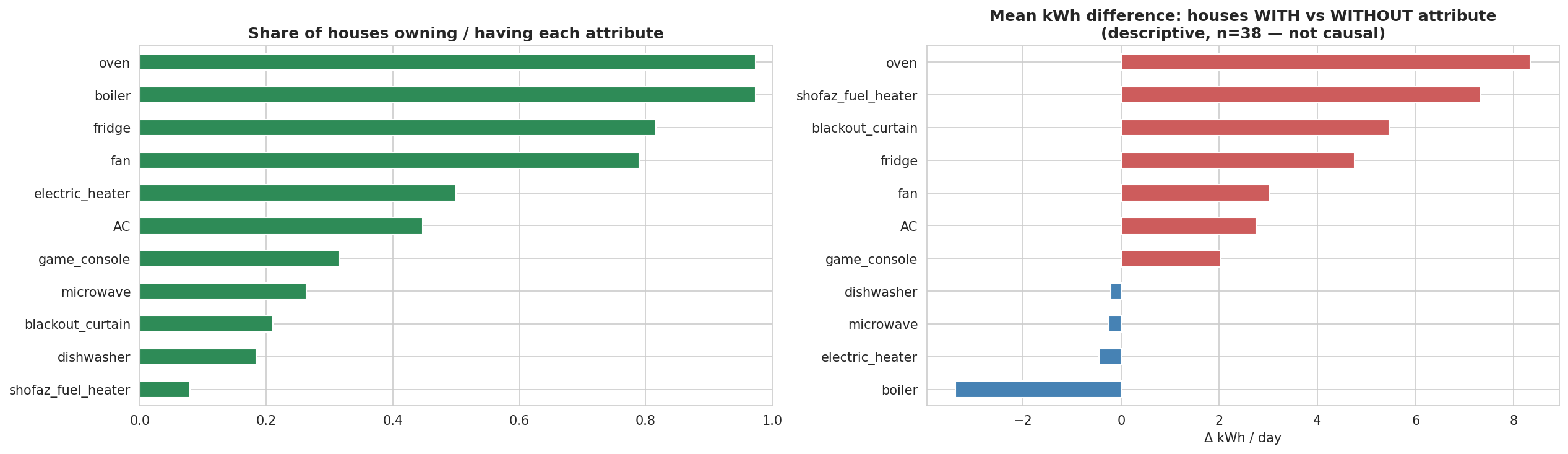}{Share of houses that own or have each binary survey attribute (left), and the difference in mean daily consumption between houses with and without it (right). With 38 houses these differences are descriptive only.}

\fig{1.0}{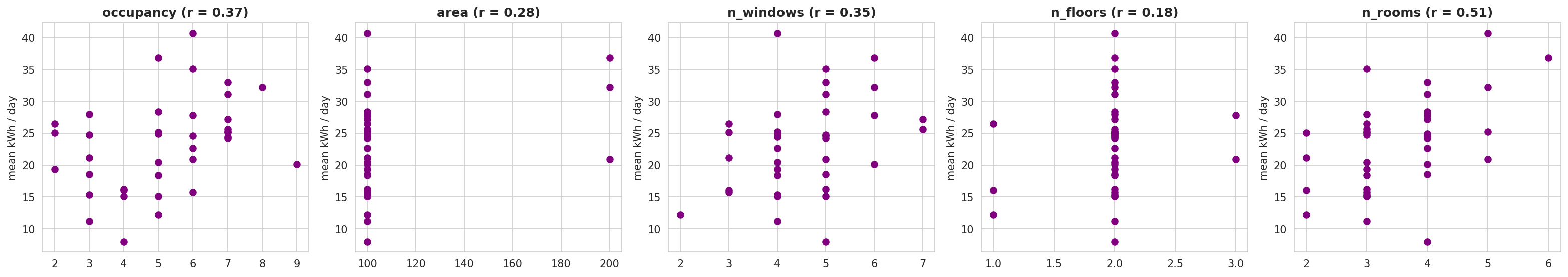}{Mean daily consumption per house against the numeric survey attributes, with Pearson correlations.}

\subsection{Cleaning and preprocessing}
The pipeline removes exact duplicate rows, keeps the first record wherever a house has two records for one date, and re-indexes every house to a complete daily calendar. None of these steps changed the data, which were already complete. It then normalises the text of categorical attributes (trimming spaces, lower-casing and correcting spelling variants), so that one category is not split in two. Constant columns and columns identical to another column are dropped. The solar energy column is rescaled, and irradiance values above 1.5\,kW/m$^2$ are replaced by interpolation over time. Near-duplicate columns ($|r|>0.98$) and near-constant columns are reported but kept. House-days below 1\,kWh are flagged as likely outages, and dates whose temperature fails the weather screen are flagged as suspect. Neither group is removed: removing them would make the test easier than real use. Their effect is examined separately in Section~\ref{sec:sens}.


\section{Methodology}\label{sec:method}

\subsection{Forecasting task}
Let $y_{i,t}$ denote the electricity consumed by house $i$ ($i=1,\dots,N$, $N=38$) on day $t$. At a forecast origin $\tau$ a model may use only information available up to and including $\tau$. For every lead time $h=1,\dots,H$ ($H=30$) it returns predictive quantiles at the levels $\alpha\in\mathcal{A}=\{0.1,0.2,\dots,0.9\}$,
\begin{equation}
\hat{q}^{(\alpha)}_{i,\tau+h\mid\tau}=F^{-1}_{i,\tau+h\mid\tau}(\alpha),
\end{equation}
where $F_{i,\tau+h\mid\tau}$ is the model's predictive distribution. The median $\hat{q}^{(0.5)}$ serves as the point forecast $\hat{y}_{i,\tau+h}$. Two information settings are compared. In the univariate track (\emph{uni}), a model sees past consumption and, if its architecture allows, the survey attributes and calendar variables. In the weather track (\emph{exog}), it additionally receives the recorded weather for the forecast days. Because weather forecasts carry error, this track was designed as an optimistic upper bound on the value of weather information. Section~\ref{sec:wxqc} shows that the recorded weather may itself be unreliable.

\subsection{Features}
Heating and cooling needs are summarised by degree days relative to comfort thresholds. With $T_t$ the daily mean air temperature,
\begin{equation}
\mathrm{HDD}_t=\max\left(0,\,18-T_t\right),\qquad \mathrm{CDD}_t=\max\left(0,\,T_t-24\right).
\end{equation}
Because buildings and occupants respond to the last few days of weather, the weather track also includes the previous day's temperature and 3- and 7-day moving averages of temperature and degree days. Calendar inputs are Friday and Saturday indicators; Iraqi public holidays, including both Eid festivals (92 dates in the record); and annual Fourier terms
\begin{equation}
s_t=\sin\!\left(\frac{2\pi d_t}{365.25}\right),\qquad c_t=\cos\!\left(\frac{2\pi d_t}{365.25}\right),
\end{equation}
where $d_t$ is the day of the year. Wind direction is an angle and enters only through its sine and cosine. After cleaning, 22 survey attributes remain; categorical ones are one-hot encoded.

\subsection{Baselines and local statistical models}
The naive forecast repeats the last observation, and the seasonal-naive forecast repeats the value from the same weekday in the most recent week:
\begin{equation}
\hat{y}^{\text{naive}}_{i,\tau+h}=y_{i,\tau},\qquad
\hat{y}^{\text{snaive}}_{i,\tau+h}=y_{i,\tau+h-7\lceil h/7\rceil}.
\end{equation}
The seasonal-naive forecast is the reference for all relative scores. Both rules produce predictive quantiles by assuming Gaussian errors, with the standard deviation $\hat{\sigma}$ estimated from each rule's in-sample residuals ($y_t-y_{t-1}$ for the naive rule and $y_t-y_{t-7}$ for the seasonal-naive rule) and widened with the horizon in the standard way~\cite{fpp3}:
\begin{equation}
\hat{\sigma}^{\text{naive}}_{h}=\hat{\sigma}\sqrt{h},\qquad
\hat{\sigma}^{\text{snaive}}_{h}=\hat{\sigma}\sqrt{\lfloor (h-1)/7\rfloor+1},\qquad
\hat{q}^{(\alpha)}_{i,\tau+h}=\hat{y}_{i,\tau+h}+z_{\alpha}\,\hat{\sigma}_{h},
\end{equation}
where $z_\alpha$ is the standard normal quantile. The random-walk width $\sqrt{h}$ explains why the naive forecast's intervals become very wide at long horizons.

AutoETS fits exponential smoothing models in state-space form~\cite{hyndman2002} and selects the error, trend and seasonal components by the corrected Akaike information criterion. As an illustration, the model with additive errors, level $\ell_t$ and weekly seasonal state $s_t$ is
\begin{equation}
\begin{aligned}
y_t&=\ell_{t-1}+s_{t-7}+\varepsilon_t,\qquad \varepsilon_t\sim\mathcal{N}(0,\sigma^2),\\
\ell_t&=\ell_{t-1}+\alpha\,\varepsilon_t,\\
s_t&=s_{t-7}+\gamma\,\varepsilon_t .
\end{aligned}
\end{equation}
AutoARIMA~\cite{hyndman2008} searches the seasonal ARIMA family with period seven,
\begin{equation}
\phi_p(B)\,\Phi_P(B^{7})\,(1-B)^{d}\,(1-B^{7})^{D}\,y_t=\theta_q(B)\,\Theta_Q(B^{7})\,\varepsilon_t ,
\end{equation}
with $B$ the backshift operator. ARIMAX adds a linear regression on heating and cooling degree days, relative humidity and the two weekend indicators, with ARIMA errors. This regressor set is restricted deliberately, because the full weather set is nearly collinear and makes the regression rank-deficient. To keep the automatic order search tractable, both ARIMA variants are fitted on the most recent 730 days at each refit. AutoTheta~\cite{theta2000} completes this family. All local models are refitted every fourth origin (every 28 days), and their states are updated with the new observations in between. Their prediction intervals follow from their Gaussian error assumptions.

\subsection{Global gradient-boosted trees}\label{sec:gbm}
LightGBM~\cite{lightgbm} and XGBoost~\cite{xgboost} are each trained as one model across all houses. A training row describes one house on one day. Its features are the house's consumption lagged by 1, 2, 3, 7, 14, 21, 28 and 364 days; rolling 7- and 28-day means and a 7-day standard deviation; a 4-week rolling mean of the 7-day lag; an expanding mean; day of week, month and day of year; the survey attributes; and, in the weather track, the weather and holiday features. Each house's series is standardised by its own mean and standard deviation before training, so that large and small consumers contribute comparably. Multi-day forecasts are produced recursively, with each prediction fed back as a lag for the next day.

Because consumption is right-skewed (Section~\ref{sec:eda}), both models minimise absolute rather than squared error. Their output therefore estimates the conditional median, which is the quantity scored as the point forecast:
\begin{equation}
\hat{f}=\arg\min_{f}\sum_{i,t}\bigl|y_{i,t}-f(\mathbf{x}_{i,t})\bigr|.
\end{equation}
An earlier run that used squared error placed 57 percent of outcomes below the reported ``median''. With the absolute-error objective this share falls to 51 percent (Section~\ref{sec:calib}).

Prediction intervals are built by split conformal prediction~\cite{vovk2005}. At each refit, the model is back-tested on the last eight 30-day windows of its own training data. For house $i$, lead time $h$ and nominal coverage $1-\beta$, the half-width $\hat{e}^{(1-\beta)}_{i,h}$ is the empirical $(1-\beta)$-quantile of the absolute back-test errors, giving
\begin{equation}
\Bigl[\hat{y}_{i,\tau+h}-\hat{e}^{(1-\beta)}_{i,h},\;\;\hat{y}_{i,\tau+h}+\hat{e}^{(1-\beta)}_{i,h}\Bigr].
\end{equation}
The models are refitted every fourth origin, and conformal half-widths are reused between refits. LightGBM's hyperparameters (number of trees, learning rate, number of leaves, minimum leaf size, row and column subsampling, L2 penalty) are chosen by 25 trials of the tree-structured Parzen estimator in Optuna~\cite{optuna}, separately for each track. The objective is the scaled absolute error of 30-day forecasts made every 14 days inside the 180-day validation window. The model is trained only on data before that window, so the test period plays no part in tuning. XGBoost uses fixed settings (500 trees, learning rate 0.03, depth 6).

\subsection{Global deep networks}
Four architectures are implemented with NeuralForecast~\cite{nixtla}: N-HiTS~\cite{nhits}, PatchTST~\cite{patchtst}, the Temporal Fusion Transformer (TFT)~\cite{tft} and DeepAR~\cite{deepar}. N-HiTS, PatchTST and TFT output the nine quantiles directly and are trained on the average pinball loss,
\begin{equation}
\mathcal{L}=\frac{1}{|\mathcal{A}|\,H}\sum_{\alpha\in\mathcal{A}}\sum_{h=1}^{H}\rho_{\alpha}\!\left(y_{\tau+h}-\hat{q}^{(\alpha)}_{\tau+h}\right),\qquad
\rho_{\alpha}(u)=u\left(\alpha-\mathbb{1}\{u<0\}\right).
\end{equation}
DeepAR maximises the likelihood of a Student-$t$ distribution whose parameters come from a two-layer recurrent network; its quantiles are taken from that distribution. All four networks receive 120 days of history, use robust scaling within each input window, and are trained once on data before the test period. Training runs for at most 1{,}500 steps, with early stopping on the validation window, and the trained networks are then applied at every test origin without retraining. N-HiTS, TFT and DeepAR also receive the survey attributes as static inputs. PatchTST accepts no external inputs and so appears only in the univariate track. Each network is trained with three random seeds, and its forecasts are averaged over seeds.

\subsection{Zero-shot foundation models}
Three pre-trained models are applied without any training or tuning on ESHED-38: Chronos-Bolt (base) and Chronos-2~\cite{chronos,chronos2}, and TimesFM-2.5 with 200 million parameters~\cite{timesfm,timesfm25}. At each origin they receive up to 512 days of each house's history and return the nine quantiles for the following 30 days. Chronos-2 accepts covariates and is therefore run in both tracks; the other two are univariate.

\subsection{Ensemble}
Combining forecasts from methods that err in different ways is one of the most dependable ways to gain accuracy~\cite{m4,m5}. The ensemble members were fixed before any test result was seen: one strong representative of each family. The univariate ensemble combines seven members (AutoETS, LightGBM, XGBoost, N-HiTS, TFT, Chronos-2 and TimesFM-2.5). The weather ensemble combines eight members: ARIMAX and the weather versions of LightGBM, XGBoost, N-HiTS, TFT and Chronos-2, plus the univariate-only AutoETS and TimesFM-2.5. Deep members are first averaged over seeds, and the quantiles are then averaged level by level (Vincentisation~\cite{genest1992}):
\begin{equation}
\hat{q}^{(\alpha)}_{\mathrm{ens}}=\frac{1}{M}\sum_{m=1}^{M}\hat{q}^{(\alpha)}_{m}.
\end{equation}
Averaging quantiles always yields a monotone set of quantiles, and it gives a sharper combined distribution than averaging the members' probability distributions.


\section{Evaluation Design}\label{sec:eval}

\subsection{Rolling-origin protocol}
Splitting days at random into training and test sets would let a model learn from days that come after the ones it is tested on. Rolling-origin evaluation is therefore used~\cite{tashman2000,bergmeir2012}. The test period runs from 24 June 2024 to 17 June 2025 (359 days). The 180 days before it serve only for early stopping and tuning, and all earlier data are for training. Inside the test period, a forecast origin is placed every seven days, giving $K=48$ origins
\begin{equation}
\tau_k=T_{\mathrm{end}}-H-7\,(K-k),\qquad k=1,\dots,48 .
\end{equation}
At each origin every configuration forecasts the next 30 days for all 38 houses, which gives $48\times30\times38=54{,}720$ forecasts per configuration (\Fig{12_rolling_origin_protocol}). Before computing any score, the evaluation code verifies that all configurations cover exactly the same house--origin--lead combinations. All predictive quantiles are truncated at zero, since consumption cannot be negative, and sorted within each forecast to remove any quantile crossing. Forecasts are scored over three lead-time buckets: h1 (the next day), h7 (days 1 to 7) and h30 (days 1 to 30). All origins fall on a Sunday.

\fig{1.0}{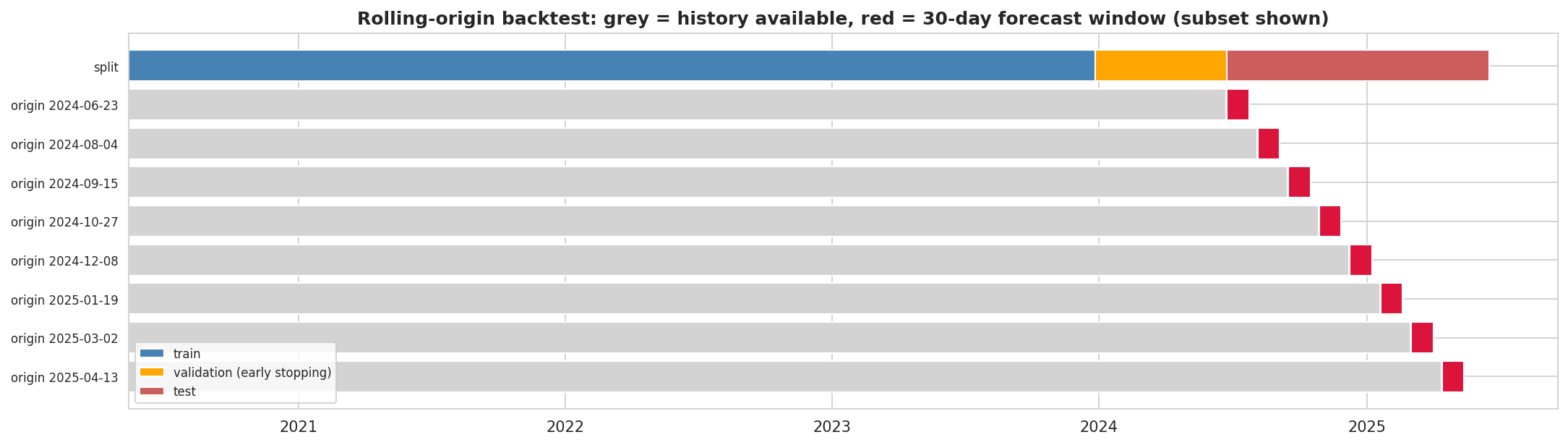}{Data split (top bar: training, validation and test) and a subset of the 48 forecast origins. Grey bars show the history available at each origin and red bars the 30-day forecast window.}

\subsection{Point accuracy}
For house $i$ and the set $S$ of origin--lead pairs in a bucket, the following are computed:
\begin{gather}
\mathrm{MAE}_i=\frac{1}{|S|}\sum_{(\tau,h)\in S}\bigl|y_{i,\tau+h}-\hat{y}_{i,\tau+h}\bigr|,\qquad
\mathrm{RMSE}_i=\Bigl(\frac{1}{|S|}\sum_{(\tau,h)\in S}\bigl(y_{i,\tau+h}-\hat{y}_{i,\tau+h}\bigr)^2\Bigr)^{1/2},\\
\mathrm{sMAPE}_i=\frac{100}{|S|}\sum_{(\tau,h)\in S}\frac{2\,\bigl|y_{i,\tau+h}-\hat{y}_{i,\tau+h}\bigr|}{|y_{i,\tau+h}|+|\hat{y}_{i,\tau+h}|}.
\end{gather}
The symmetric percentage error is used because the ordinary percentage error is undefined on zero-consumption days. Because houses differ fivefold in scale, the main point metric is the mean absolute scaled error~\cite{hyndman2006}. It divides the MAE by the in-sample error of the weekly seasonal-naive forecast over the $T_0$ days that precede the test period (14 May 2020 to 23 June 2024, i.e.\ the training and validation data):
\begin{equation}
\mathrm{MASE}_i=\frac{\mathrm{MAE}_i}{\frac{1}{T_0-7}\sum_{t=8}^{T_0}\bigl|y_{i,t}-y_{i,t-7}\bigr|}.
\end{equation}
To compare configurations on the same test days, the geometric mean over houses of each configuration's MASE relative to that of the seasonal-naive forecast is also reported,
\begin{equation}
\mathrm{rel\text{-}gMASE}_m=\Bigl(\prod_{i=1}^{N}\frac{\mathrm{MASE}_{i,m}}{\mathrm{MASE}_{i,\mathrm{snaive}}}\Bigr)^{1/N},
\end{equation}
which keeps a few difficult houses from dominating the summary. A pooled coefficient of determination over all $|\mathcal{F}|$ forecasts in a bucket completes the point metrics,
\begin{equation}
R^2=1-\frac{\sum_{(i,\tau,h)\in\mathcal{F}}\bigl(y_{i,\tau+h}-\hat{y}_{i,\tau+h}\bigr)^2}{\sum_{(i,\tau,h)\in\mathcal{F}}\bigl(y_{i,\tau+h}-\bar{y}\bigr)^2},
\end{equation}
where $\bar{y}$ is the mean of the observed values in $\mathcal{F}$.

\subsection{Probabilistic quality}
The continuous ranked probability score (CRPS) is a strictly proper score for a whole predictive distribution~\cite{gneiting2007a}. It is approximated from the nine quantiles through the pinball loss,
\begin{equation}
\mathrm{CRPS}(F,y)=\int_0^1 2\,\rho_\alpha\!\left(y-F^{-1}(\alpha)\right)\mathrm{d}\alpha
\;\approx\;\frac{2}{|\mathcal{A}|}\sum_{\alpha\in\mathcal{A}}\rho_\alpha\!\left(y-\hat{q}^{(\alpha)}\right),
\end{equation}
and divided by the same in-sample scale as MASE to obtain a scaled CRPS (sCRPS). Because only the levels 0.1 to 0.9 are available, this approximation omits the outer tails and slightly understates the CRPS, equally for every configuration. The relative score reported alongside rel-gMASE is the ratio of house-averaged sCRPS values,
\begin{equation}
\mathrm{rel\text{-}sCRPS}_m=\frac{\frac{1}{N}\sum_{i=1}^{N}\mathrm{sCRPS}_{i,m}}{\frac{1}{N}\sum_{i=1}^{N}\mathrm{sCRPS}_{i,\mathrm{snaive}}}.
\end{equation}
Following Gneiting et al.~\cite{gneiting2007b}, calibration and sharpness are reported separately. Calibration is measured by the empirical coverage of the central 80 percent interval and by reliability curves of observed against nominal quantile levels. Sharpness is measured by the mean interval width:
\begin{equation}
\mathrm{Cov}_{80}=\frac{1}{|S|}\sum_{(\tau,h)\in S}\mathbb{1}\bigl\{\hat{q}^{(0.1)}\le y_{\tau+h}\le \hat{q}^{(0.9)}\bigr\},\qquad
W_{80}=\frac{1}{|S|}\sum_{(\tau,h)\in S}\bigl(\hat{q}^{(0.9)}-\hat{q}^{(0.1)}\bigr).
\end{equation}

\subsection{Statistical tests}
Each house is treated as a paired observation. Following Dem\v{s}ar~\cite{demsar2006}, the $k=24$ configurations are ranked within each house by sCRPS and compared with the Friedman test~\cite{friedman1937},
\begin{equation}
\chi^2_F=\frac{12N}{k(k+1)}\Bigl[\sum_{j=1}^{k}\bar{R}_j^{\,2}-\frac{k(k+1)^2}{4}\Bigr],
\end{equation}
where $\bar{R}_j$ is the mean rank of configuration $j$. When the test rejects, the Nemenyi procedure~\cite{nemenyi1963} declares two configurations different if their mean ranks differ by more than
\begin{equation}
\mathrm{CD}=q_{\alpha}\sqrt{\frac{k(k+1)}{6N}},
\end{equation}
with $q_\alpha$ derived from the studentised range distribution. Nemenyi is conservative with many configurations, so the best-ranked configuration is also compared with every other one using the Wilcoxon signed-rank test~\cite{wilcoxon1945} on per-house sCRPS. The resulting $m$ $p$-values are adjusted with Holm's procedure~\cite{holm1979},
\begin{equation}
\tilde{p}_{(j)}=\max_{l\le j}\;\min\bigl\{1,\;(m-l+1)\,p_{(l)}\bigr\},
\end{equation}
where $p_{(l)}$ is the $l$-th smallest raw $p$-value. All tests use $\alpha=0.05$.

\subsection{Implementation}
The pipeline is written in Python. Local models use StatsForecast, gradient-boosted trees MLForecast with LightGBM and XGBoost, and deep networks NeuralForecast~\cite{nixtla}. Foundation models come from the \texttt{chronos-forecasting} and \texttt{timesfm} packages, and tuning uses Optuna. All experiments ran in a Kaggle notebook with one NVIDIA GPU; every random seed is fixed.


\section{Results}\label{sec:results}

\subsection{Overall accuracy at a one-week horizon}\label{sec:overall}
Table~\ref{tab:h7} reports every configuration at the h7 horizon, ordered by sCRPS. The leading configurations bring MASE down from 0.861 for the seasonal-naive forecast to 0.673--0.674, cut MAE from 7.05 to about 5.52\,kWh per day, and raise the pooled $R^2$ from 0.41 to 0.64. \Fig{13_leaderboard_bars} shows the same comparison for all three horizons, coloured by family.

\begin{table}[htbp]
\centering\footnotesize
\setlength{\tabcolsep}{3pt}
\caption{Accuracy at the h7 horizon (days 1--7), averaged over 38 houses and 48 origins. Rank is the mean rank of per-house sCRPS among 24 configurations (lower is better). Deep models are averaged over three seeds; their seed standard deviation in MASE is at most 0.016. Best values in bold.}\label{tab:h7}
\begin{tabular}{llrrrrrrrr}
\toprule
Configuration & Family & MAE & sMAPE & MASE & $R^2$ & sCRPS & Cov$_{80}$ & $W_{80}$ & Rank\\
 & & (kWh) & (\%) & & & & & (kWh) & \\
\midrule
Chronos-2 [uni] & Foundation & 5.53 & 30.1 & \textbf{0.673} & 0.636 & \textbf{0.528} & 0.78 & 16.7 & 3.74\\
TimesFM-2.5 [uni] & Foundation & \textbf{5.52} & \textbf{30.0} & 0.674 & 0.639 & 0.529 & 0.81 & 17.6 & 3.87\\
Ensemble [uni] & Ensemble & 5.53 & \textbf{30.0} & 0.674 & \textbf{0.646} & 0.531 & 0.81 & 17.8 & \textbf{3.55}\\
Chronos-Bolt [uni] & Foundation & 5.58 & 30.3 & 0.680 & 0.632 & 0.533 & 0.80 & 17.5 & 5.13\\
Ensemble [exog] & Ensemble & 5.57 & 30.2 & 0.679 & 0.643 & 0.534 & 0.80 & 17.5 & 4.95\\
Chronos-2 [exog] & Foundation & 5.65 & 30.6 & 0.689 & 0.619 & 0.539 & 0.78 & 17.2 & 7.97\\
AutoETS [uni] & Local & 5.60 & 30.3 & 0.683 & 0.637 & 0.543 & 0.82 & 18.9 & 8.05\\
N-HiTS [uni] & Deep & 5.66 & 30.8 & 0.691 & 0.629 & 0.545 & 0.82 & 19.2 & 9.11\\
AutoARIMA [uni] & Local & 5.60 & 30.5 & 0.683 & 0.636 & 0.546 & 0.82 & 18.7 & 9.42\\
ARIMAX [exog] & Local & 5.67 & 30.9 & 0.691 & 0.631 & 0.550 & 0.81 & 18.5 & 10.05\\
AutoTheta [uni] & Local & 5.59 & 30.3 & 0.682 & 0.638 & 0.551 & 0.84 & 19.6 & 11.97\\
DeepAR [uni] & Deep & 5.76 & 31.1 & 0.702 & 0.619 & 0.553 & 0.75 & 16.6 & 11.18\\
PatchTST [uni] & Deep & 5.82 & 31.5 & 0.709 & 0.611 & 0.560 & 0.76 & 17.2 & 13.61\\
DeepAR [exog] & Deep & 5.84 & 31.5 & 0.713 & 0.608 & 0.563 & 0.75 & 17.1 & 14.50\\
XGBoost [uni] & Tree ensemble & 5.65 & 30.6 & 0.688 & 0.636 & 0.564 & 0.74 & 16.7 & 14.63\\
TFT [uni] & Deep & 5.92 & 31.9 & 0.722 & 0.602 & 0.564 & 0.79 & 18.6 & 14.42\\
LightGBM [uni] & Tree ensemble & 5.74 & 30.9 & 0.698 & 0.623 & 0.573 & 0.73 & 16.7 & 15.82\\
XGBoost [exog] & Tree ensemble & 5.75 & 31.2 & 0.701 & 0.627 & 0.574 & 0.73 & 16.8 & 17.97\\
LightGBM, no survey [exog] & Tree ensemble & 5.76 & 30.9 & 0.700 & 0.622 & 0.574 & 0.73 & 16.7 & 15.92\\
LightGBM [exog] & Tree ensemble & 5.76 & 30.9 & 0.701 & 0.622 & 0.574 & 0.73 & 16.7 & 15.97\\
N-HiTS [exog] & Deep & 6.31 & 33.9 & 0.770 & 0.550 & 0.605 & 0.72 & 17.6 & 20.29\\
TFT [exog] & Deep & 6.40 & 33.6 & 0.781 & 0.530 & 0.616 & 0.71 & 16.6 & 21.11\\
Seasonal naive [uni] & Baseline & 7.05 & 38.2 & 0.861 & 0.414 & 0.696 & 0.86 & 25.7 & 22.76\\
Naive [uni] & Baseline & 7.04 & 37.7 & 0.859 & 0.385 & 0.814 & 0.96 & 42.6 & 24.00\\
\bottomrule
\end{tabular}
\end{table}

\fig{1.0}{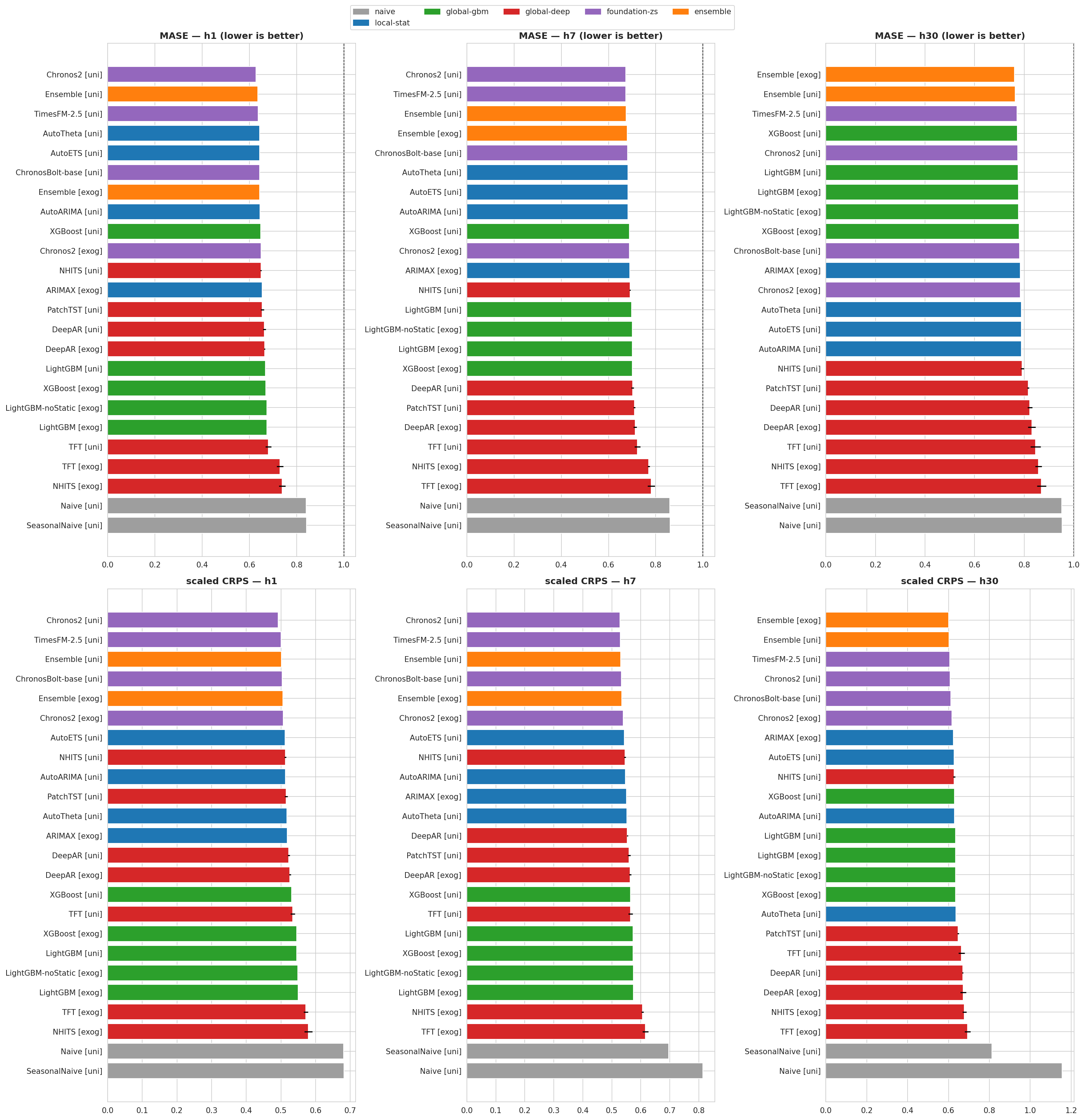}{MASE (top row) and sCRPS (bottom row) of all 24 configurations at h1, h7 and h30, coloured by family (grey: naive; blue: local statistical; green: tree ensembles; red: deep; purple: foundation; orange: ensemble). Error bars give the seed standard deviation of deep models; the dashed line marks MASE~=~1.}

Three features of these results stand out. First, the seasonal-naive forecast has a MASE below one even though MASE is scaled by its own error. The reason is that the scale is computed on all data before the test period (May 2020 to June 2024), which was more volatile than the test period, so every model looks somewhat better in MASE than it would on a more typical year. The relative scores used below compare all configurations on the same test days and are not affected by this.

Second, the top of the table is crowded. Fourteen configurations, from four of the five base families plus both ensembles, lie within 0.23\,kWh of the best MAE. Differences in MAE alone therefore say little, and the probabilistic scores and the tests in Section~\ref{sec:signif} do more to separate the models.

Third, the tree ensembles rank higher on point accuracy than on sCRPS. XGBoost [uni] has a lower MAE (5.65\,kWh) than any deep network, yet is only 15th of 24 by average sCRPS, because its 80 percent intervals cover only 74 percent of outcomes (Section~\ref{sec:calib}).

\subsection{Accuracy across horizons}
Table~\ref{tab:horizons} and \Fig{14_relative_skill_heatmap} give skill relative to the seasonal-naive forecast at each horizon. One day ahead, Chronos-2 is best on both metrics: it reduces the geometric-mean error by 25.4 percent and the sCRPS by 27.8 percent. At one week the leading group is almost level, with Chronos-2, TimesFM-2.5 and the univariate ensemble each at 0.780. At 30 days the two ensembles lead, with the weather ensemble at 0.797 and 0.739. The family ordering hardly changes with horizon. The naive forecast is the exception: its sCRPS worsens from 0.999 to 1.424 of the seasonal-naive value, because random-walk intervals keep widening with the horizon. Its 80 percent interval covers 99 percent of outcomes at h30, so it is far too wide to be useful.

\begin{table}[htbp]
\centering\small
\caption{Skill relative to the seasonal-naive forecast (= 1.000) by horizon; lower is better. Selected configurations; \Fig{14_relative_skill_heatmap} shows all 24. Best values in bold.}\label{tab:horizons}
\begin{tabular}{lrrrrrr}
\toprule
& \multicolumn{3}{c}{rel-gMASE} & \multicolumn{3}{c}{rel-sCRPS}\\
\cmidrule(lr){2-4}\cmidrule(lr){5-7}
Configuration & h1 & h7 & h30 & h1 & h7 & h30\\
\midrule
Chronos-2 [uni] & \textbf{0.746} & \textbf{0.780} & 0.810 & \textbf{0.722} & \textbf{0.759} & 0.749\\
TimesFM-2.5 [uni] & 0.755 & \textbf{0.780} & 0.807 & 0.734 & 0.760 & 0.746\\
Ensemble [uni] & 0.754 & \textbf{0.780} & 0.799 & 0.735 & 0.763 & 0.742\\
Ensemble [exog] & 0.763 & 0.786 & \textbf{0.797} & 0.742 & 0.768 & \textbf{0.739}\\
Chronos-Bolt [uni] & 0.764 & 0.787 & 0.816 & 0.739 & 0.766 & 0.754\\
AutoETS [uni] & 0.763 & 0.791 & 0.824 & 0.751 & 0.780 & 0.772\\
AutoARIMA [uni] & 0.765 & 0.790 & 0.825 & 0.753 & 0.785 & 0.776\\
N-HiTS [uni] & 0.770 & 0.800 & 0.828 & 0.752 & 0.783 & 0.773\\
XGBoost [uni] & 0.768 & 0.796 & 0.809 & 0.779 & 0.811 & 0.776\\
LightGBM [uni] & 0.792 & 0.806 & 0.808 & 0.801 & 0.823 & 0.780\\
TFT [uni] & 0.804 & 0.834 & 0.880 & 0.784 & 0.811 & 0.817\\
Naive [uni] & 0.994 & 0.991 & 0.997 & 0.999 & 1.170 & 1.424\\
\bottomrule
\end{tabular}
\end{table}

\fig{0.82}{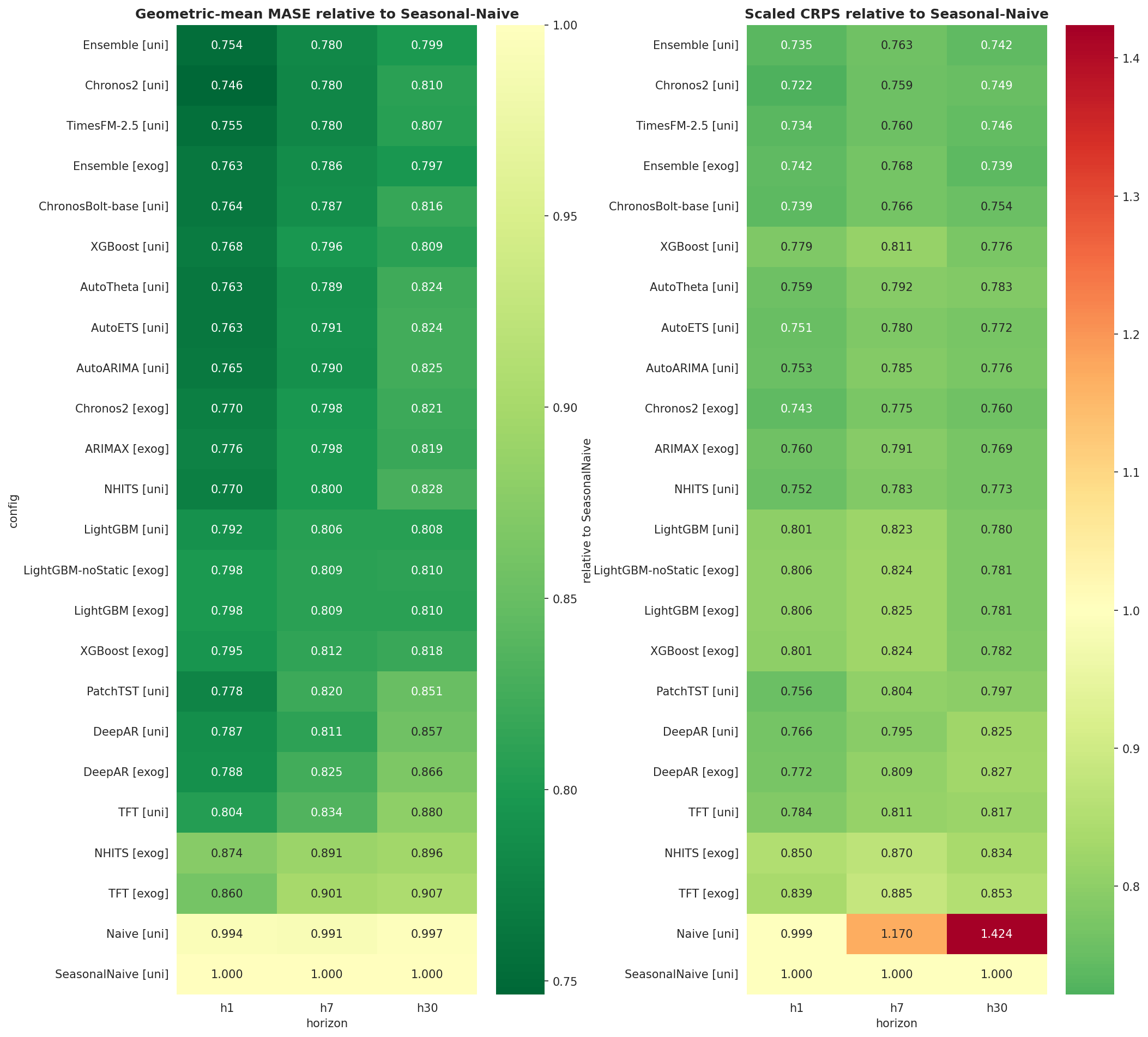}{Geometric-mean MASE (left) and scaled CRPS (right) of every configuration relative to the seasonal-naive forecast, by horizon. Green cells mark a large improvement over the baseline.}

The tree ensembles lose less ground than the other families as the horizon grows. LightGBM's relative error is 0.792 at h1 and only 0.808 at h30. At that horizon only TimesFM-2.5 is better among single models. The deep networks lose the most: TFT's relative error rises from 0.804 to 0.880. Reading the heatmap column by column, the foundation models and ensembles form a consistently dark-green block at the top, while the deep models with weather inputs and the two naive rules stay pale at every horizon.

Error also changes with lead time in a regular weekly pattern (\Fig{15_lead_time_curves}). For the univariate ensemble, MASE rises from 0.64 on day 1 to 0.83 on day 30, but not smoothly: it peaks at lead times 5, 12, 19 and 26. Every origin is a Sunday, so these lead times are all Fridays. On Fridays the ensemble's MASE averages 0.81, against 0.76 on other days and 0.72 on Tuesdays. Weekend consumption is both higher and less regular, and every model finds it harder to predict. The coverage panel of the same figure shows how well each family keeps its intervals honest as the horizon grows. The foundation models and the ensemble stay between 0.73 and 0.84 across all lead days. The tree ensembles sit lower, between 0.68 and 0.76. DeepAR with weather inputs falls from 0.78 on day 1 to 0.52 on day 30, because its intervals do not widen as uncertainty grows.

\fig{1.0}{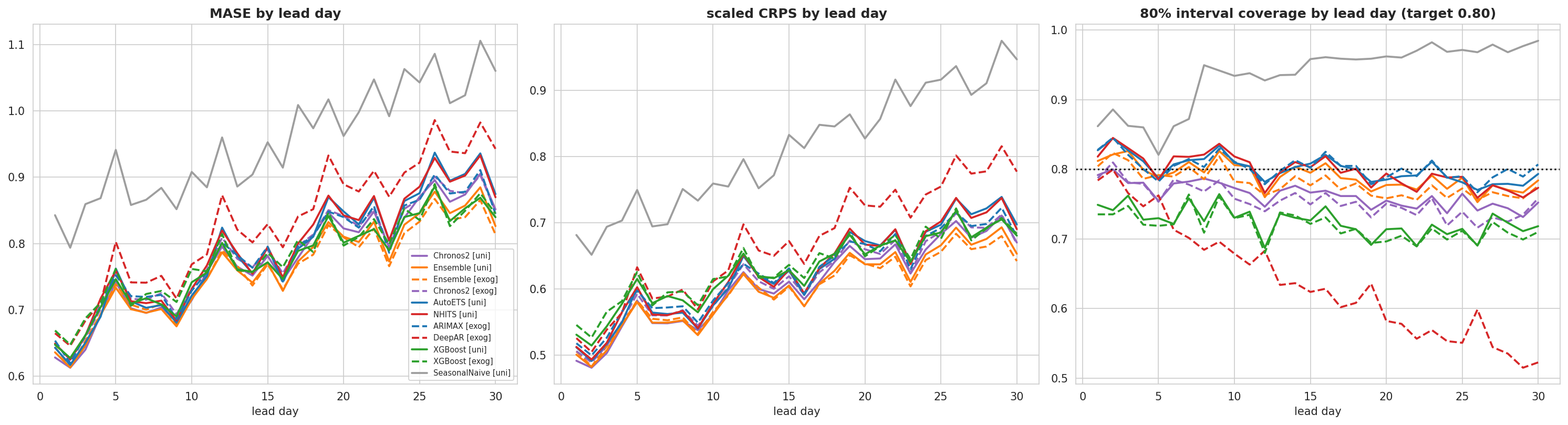}{MASE (left), scaled CRPS (centre) and 80 percent interval coverage (right) by lead day, for the best configuration of each family and the seasonal-naive baseline. Solid lines: univariate track; dashed lines: weather track. The periodic peaks at lead days 5, 12, 19 and 26 are Fridays.}

\subsection{Results by house}
The leading configurations improve on the seasonal-naive forecast in all 38 houses. Eighteen of the 23 configurations other than the baseline itself do so in every house; the exceptions are PatchTST (37 houses), TFT [uni] and N-HiTS [exog] (35 each), TFT [exog] (32), and the naive forecast. \Fig{16_per_house_mase_boxplot} shows the spread of per-house MASE. For the univariate ensemble it ranges from 0.44 (H12) to 0.96 (H35), with a median of 0.65, and every leading configuration finds the same houses easy and the same houses hard. The boxes of the foundation models, ensembles and local models are almost the same height, so they are about equally consistent across houses. The deep models with weather inputs show visibly longer upper whiskers.

\fig{1.0}{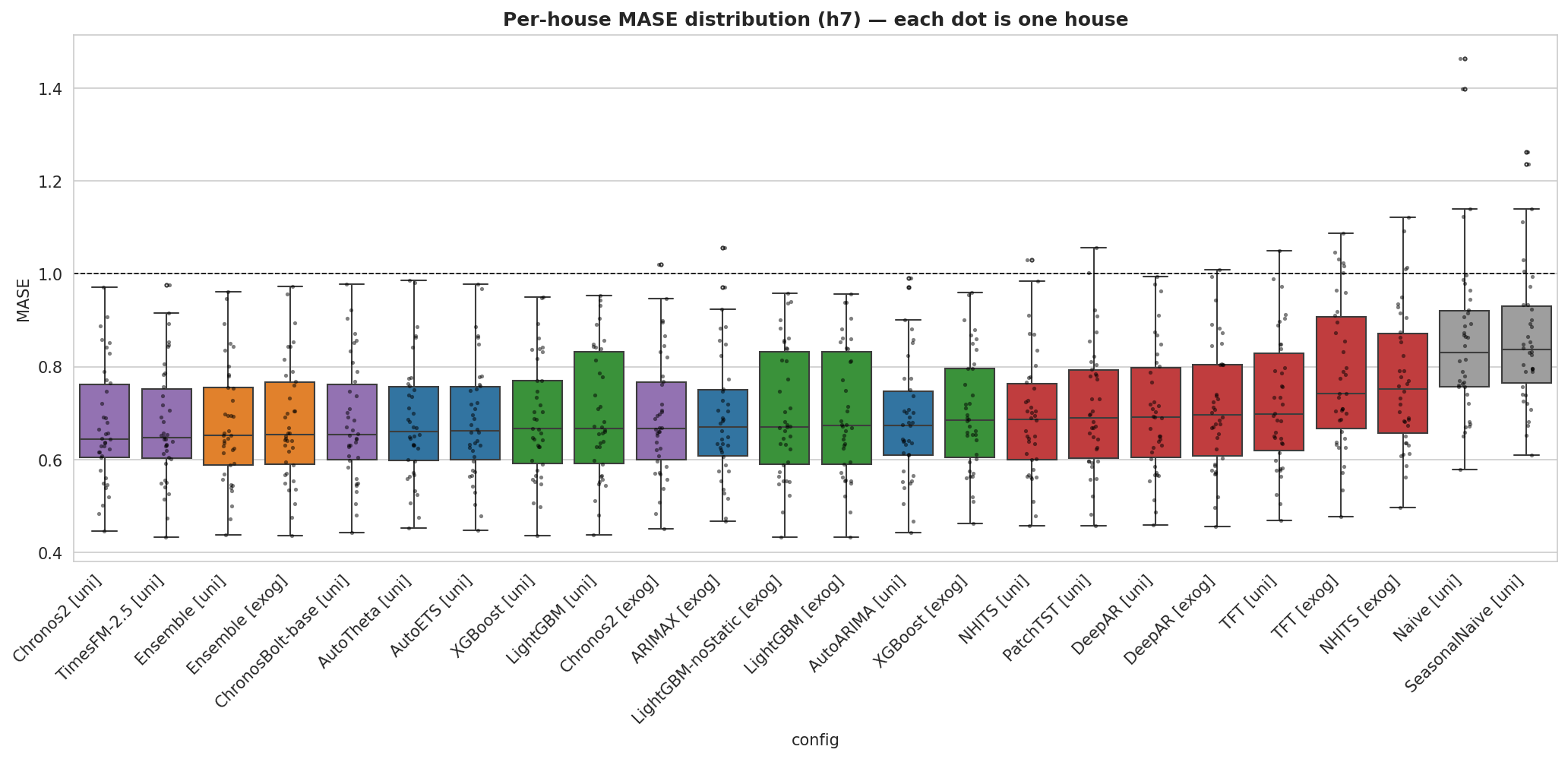}{Per-house MASE at h7 for every configuration; each dot is one house. Configurations are ordered by median MASE. The dashed line marks MASE~=~1.}

\Fig{23_pairwise_winrate} turns these per-house results into head-to-head comparisons. Each cell gives the share of houses in which the row configuration has a lower sCRPS than the column configuration. Chronos-2 [uni] beats LightGBM [uni] in 97 percent of houses and AutoETS in 84 percent. The univariate ensemble beats AutoETS in 84 percent of houses but Chronos-2 in only 39 percent. The near-solid red upper-right triangle shows how consistent the overall ordering is from house to house.

\fig{0.82}{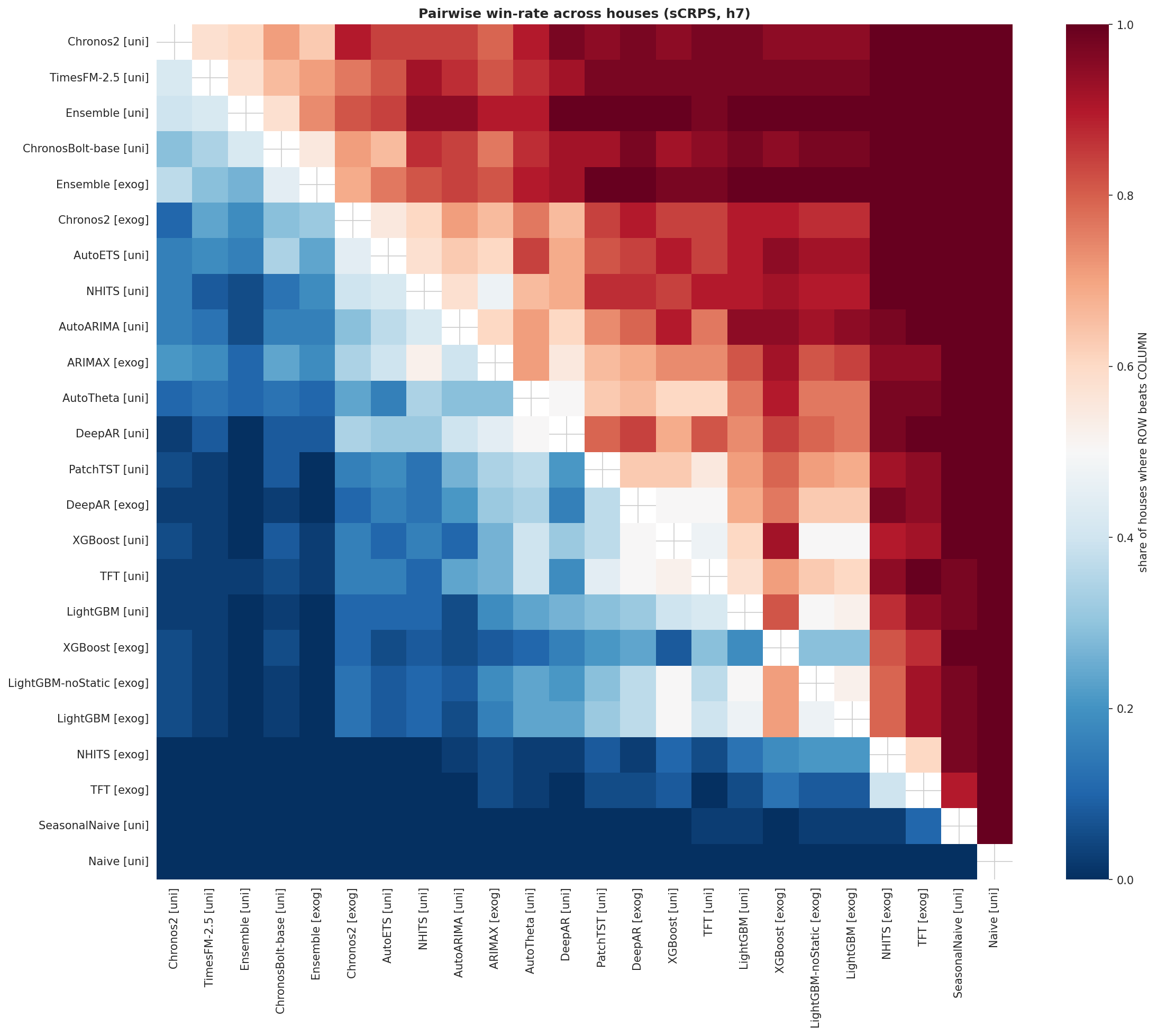}{Pairwise win rate at h7: the share of the 38 houses in which the row configuration achieves a lower sCRPS than the column configuration.}

Does pooling information across houses help? \Fig{26_global_vs_local_per_house} compares the best pooled model, N-HiTS [uni], with the best local model, AutoETS, house by house. The points lie on both sides of the diagonal: N-HiTS wins in 42 percent of houses. The right-hand panel shows that the difference is not related to how volatile a house is. LightGBM and XGBoost do worse still, beating AutoETS in only 11 percent of houses. With 38 houses and five years of data, learning from other houses does not beat a well-specified model for each house. The zero-shot foundation models, which draw on a far larger and unrelated pool of series, do beat AutoETS in most houses.

\fig{1.0}{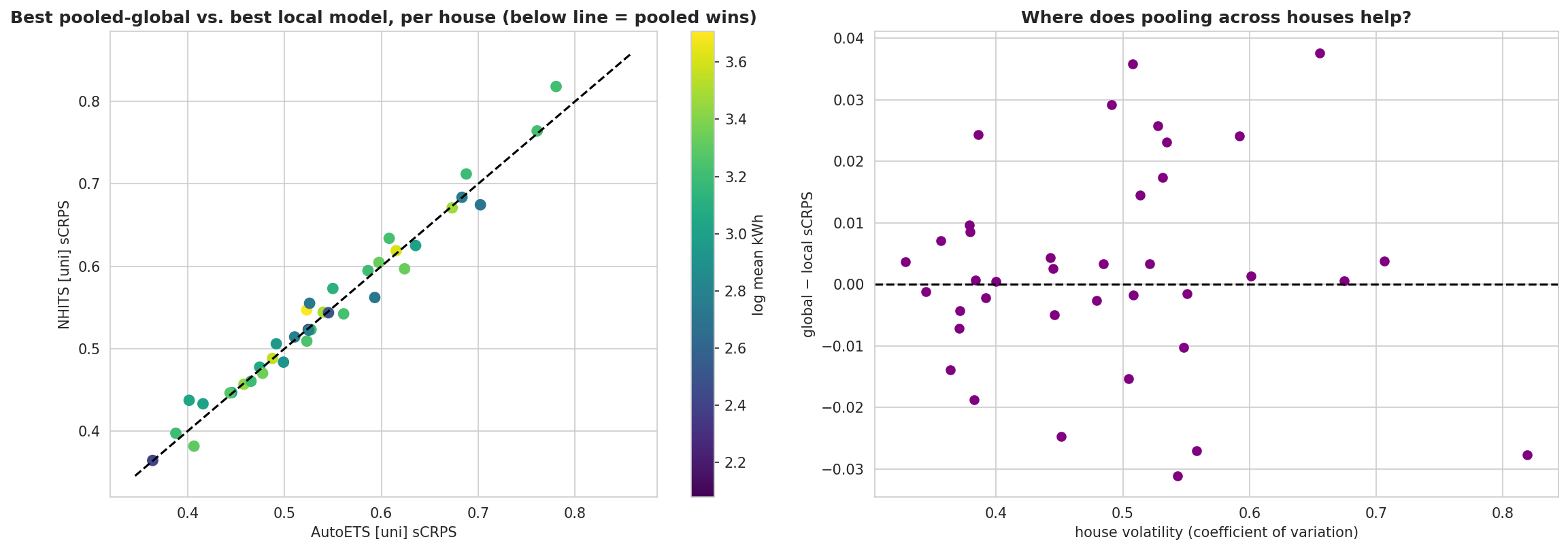}{Left: per-house sCRPS (h7) of the best pooled global model (N-HiTS [uni]) against the best local model (AutoETS [uni]), coloured by log mean consumption; points below the diagonal favour pooling. Right: the difference (global minus local) against house volatility, measured by the coefficient of variation.}

\subsection{Seasonal and temperature effects}
Forecast difficulty follows the seasons closely. \Fig{20_error_by_month} breaks down h7 error by the calendar month of the target day. For the univariate ensemble, MASE is lowest in August (0.41) and stays between 0.53 and 0.62 from April to October. It rises to 0.69 in November and to 0.79--0.91 from December to March. Every configuration, the seasonal-naive forecast included, follows the same pattern. The seasonal-naive MASE exceeds 1.0 from December to March, and the leading models improve on it most, in relative terms, in exactly those months.

\fig{0.78}{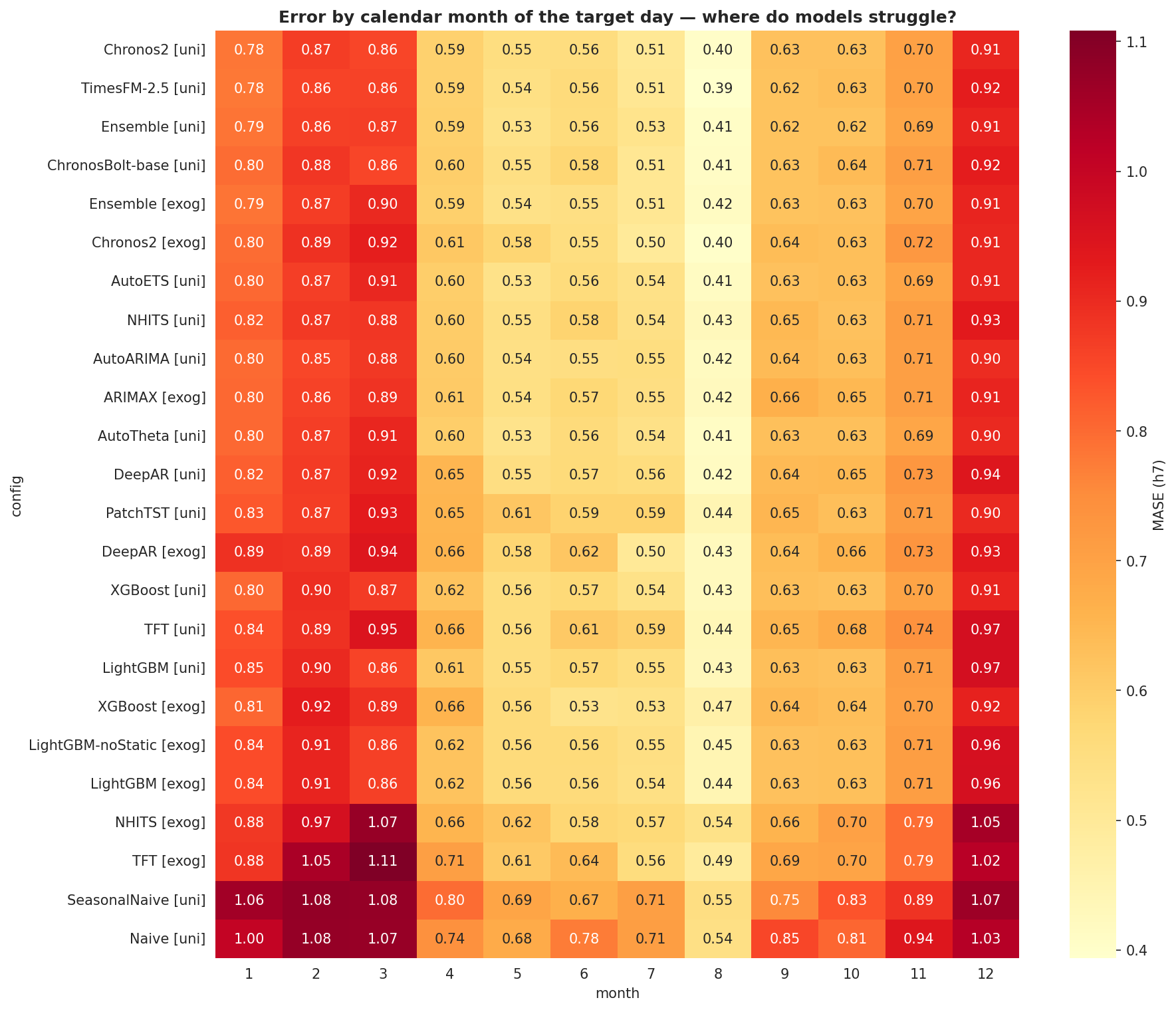}{MASE (h7) by calendar month of the target day for every configuration. Darker cells mark larger errors.}

\Fig{21_error_vs_temperature} gives the same picture in terms of temperature. The ensemble's MASE is about 0.84 on target days with a mean temperature below 15\,\degC and falls steadily to about 0.48 above 35\,\degC; the seasonal-naive forecast falls from about 1.08 to 0.62. Cold days are therefore the hardest to forecast. This is consistent with the heating-driven demand of Section~\ref{sec:eda}: switching heaters on and off is a household decision that varies from day to day, while the low summer load is more stable.

\fig{0.85}{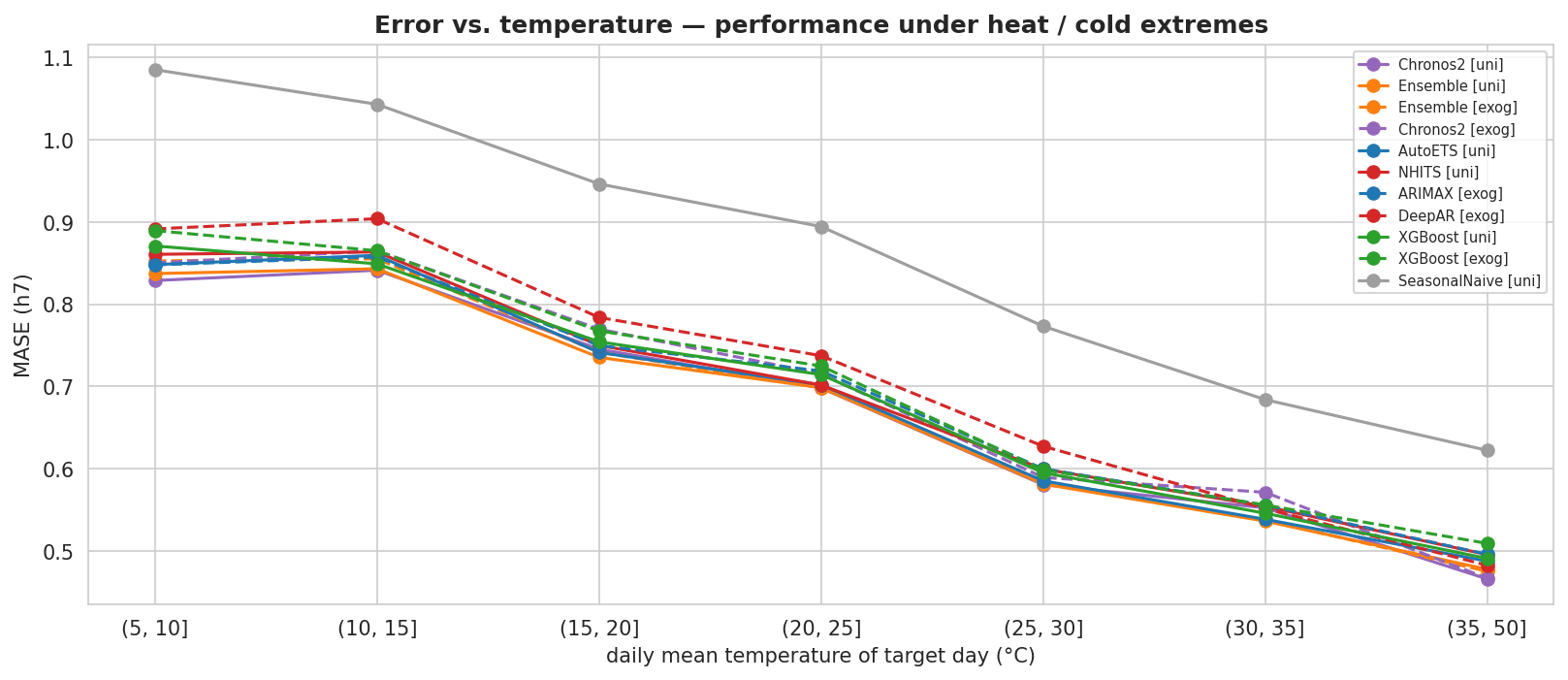}{MASE (h7) by mean temperature of the target day, for the best configuration of each family and the seasonal-naive baseline.}

\Fig{25_error_per_origin} follows the error across the 48 origins of the test period. Error is lowest in late summer, reaching 0.34 for the origin of 18 August 2024. It rises through autumn and peaks sharply for the origin of 1 December 2024 (MASE 1.17), only one week after one of the easiest origins (0.49 on 24 November). This jump marks the abrupt start of the heating season, and every model missed it, including those that were given the recorded temperature. A second, smaller peak follows on 23 February 2025 (1.09). The seasonal-naive forecast has its worst origin a week later (1.32 on 2 March), when consumption fell back as the weather warmed. Shocks of this kind are where forecasts matter most to an operator, and none of the methods tested anticipates them.

\fig{1.0}{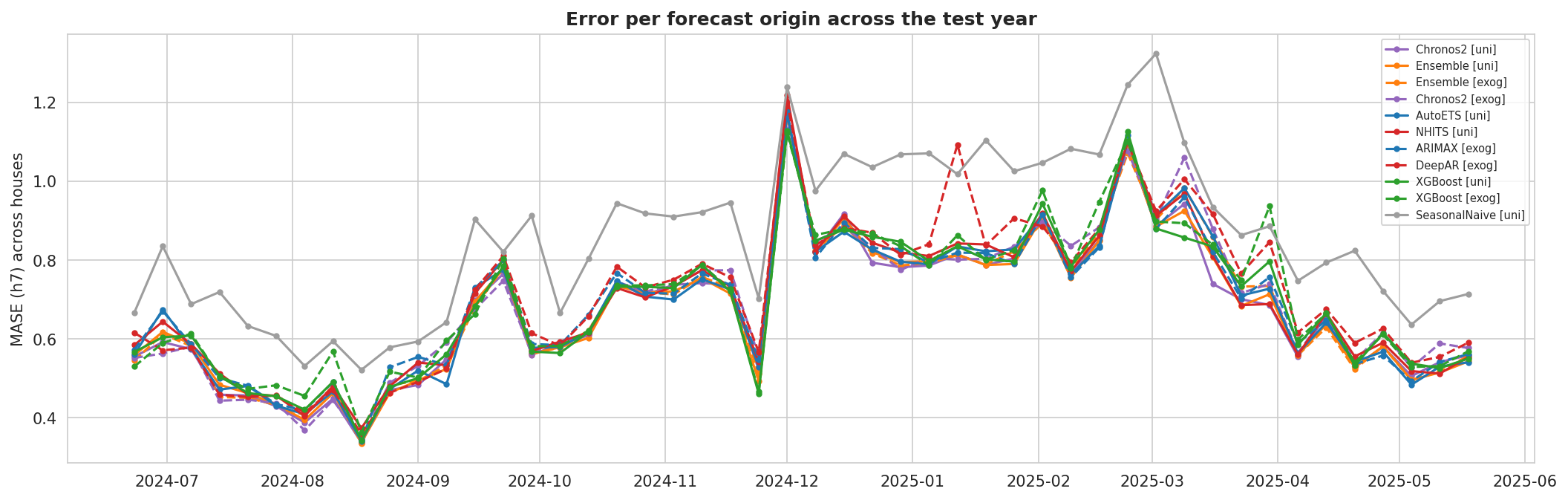}{MASE (h7) averaged over houses for each of the 48 forecast origins in the test period.}

\subsection{Calibration and sharpness}\label{sec:calib}
\Fig{19_calibration} and Table~\ref{tab:reliab} show how reliable the predictive quantiles are. A perfectly calibrated forecaster would have each nominal quantile exceeded at exactly its stated rate. The foundation models and the univariate ensemble come close: their 10th, 50th and 90th percentiles are exceeded from below at rates within 0.02 of the nominal level. The local statistical models are slightly too wide in the lower tail and well calibrated at the median. The tree ensembles are well centred now that they predict the median (0.51 at $\tau=0.5$, compared with 0.57 in an earlier run trained on squared error), but their tails are too narrow, so their 80 percent intervals cover only 73--74 percent of outcomes. The weather-track deep networks are the least reliable. Their quantile curves are flatter than the diagonal, so their intervals are too narrow, and TFT [exog] covers only 71 percent at h7. The right panel of \Fig{19_calibration} shows the trade-off between interval width and coverage. The foundation models and ensembles sit near the 0.80 line with mean widths of 16.7--17.8\,kWh. The local models are slightly wider and over-cover. The tree ensembles are as narrow as the best models, but at the cost of coverage.

\begin{table}[htbp]
\centering\small
\caption{Quantile reliability at h7: observed share of outcomes at or below each predictive quantile (ideal values 0.10, 0.50 and 0.90), and 80 percent interval coverage (ideal 0.80).}\label{tab:reliab}
\begin{tabular}{lrrrr}
\toprule
Configuration & $\tau=0.1$ & $\tau=0.5$ & $\tau=0.9$ & Cov$_{80}$\\
\midrule
Chronos-2 [uni] & 0.107 & 0.488 & 0.888 & 0.78\\
TimesFM-2.5 [uni] & 0.087 & 0.485 & 0.899 & 0.81\\
Ensemble [uni] & 0.092 & 0.505 & 0.900 & 0.81\\
AutoETS [uni] & 0.084 & 0.511 & 0.900 & 0.82\\
AutoTheta [uni] & 0.072 & 0.513 & 0.911 & 0.84\\
LightGBM [uni] & 0.119 & 0.506 & 0.853 & 0.73\\
XGBoost [uni] & 0.130 & 0.528 & 0.871 & 0.74\\
DeepAR [exog] & 0.119 & 0.514 & 0.872 & 0.75\\
TFT [exog] & 0.159 & 0.538 & 0.864 & 0.71\\
Seasonal naive [uni] & 0.068 & 0.498 & 0.928 & 0.86\\
\bottomrule
\end{tabular}
\end{table}

\fig{1.0}{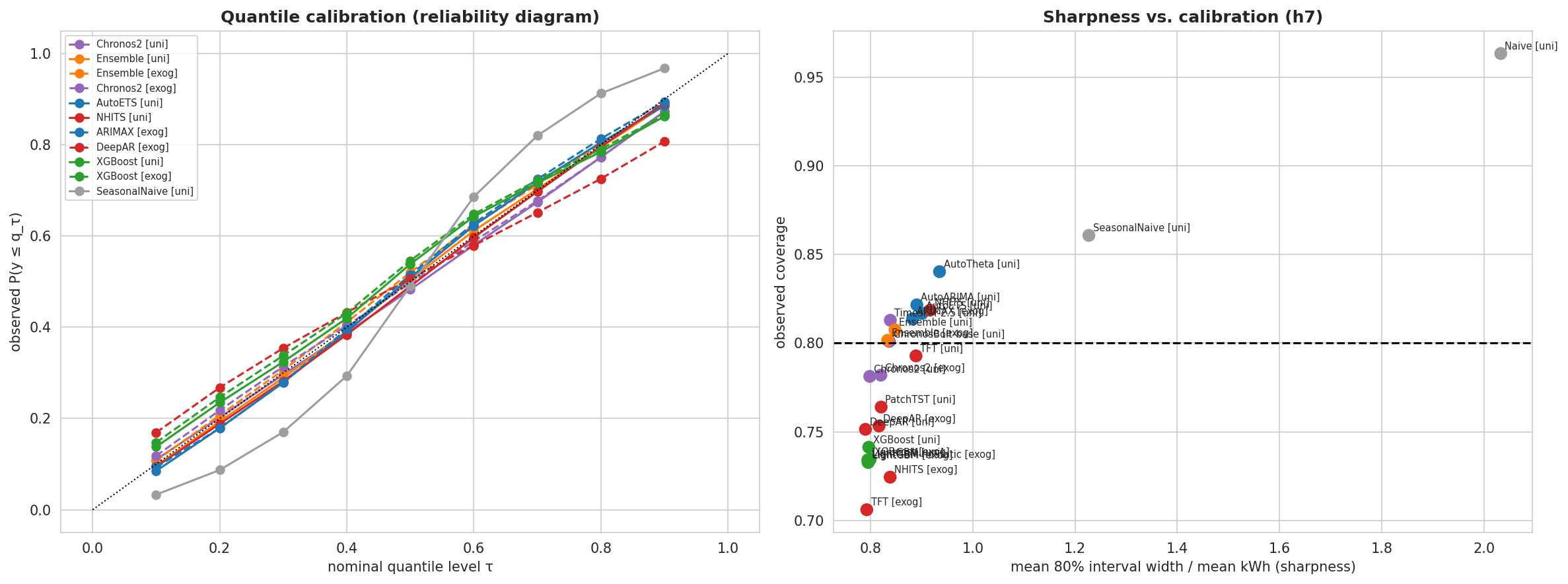}{Left: reliability diagram for the best configuration of each family, plotting the observed share of outcomes below each quantile against its nominal level; the dotted diagonal is perfect calibration. Right: mean 80 percent interval width (relative to mean consumption) against observed coverage at h7; the dashed line is the 0.80 target.}

\subsection{Statistical significance}\label{sec:signif}
The Friedman test rejects equal performance at every horizon ($\chi^2_F=451.2$, 655.5 and 539.2 for h1, h7 and h30; all $p<10^{-80}$). With 24 configurations and 38 houses, the Nemenyi critical difference is 5.90 ranks. \Fig{27_cd_diagram_h1}, \Fig{27_cd_diagram_h7} and \Fig{27_cd_diagram_h30} show the critical-difference diagrams. In each, configurations are placed by mean rank and joined by a bar when Nemenyi cannot separate them. The Holm-corrected Wilcoxon tests are more powerful and sharpen the picture:
\begin{itemize}
\item \textbf{h1:} Chronos-2 [uni] has the best mean rank (4.66) and is significantly better than all 23 other configurations, including the univariate ensemble (mean rank 5.39).
\item \textbf{h7:} the univariate ensemble has the best mean rank (3.55), but it is statistically tied with Chronos-2 (3.74), TimesFM-2.5 (3.87) and Chronos-Bolt (5.13) (adjusted $p=0.51$ for each comparison). It is significantly better than all other configurations, including the weather ensemble (4.95).
\item \textbf{h30:} the weather ensemble has the best mean rank (3.58) and is statistically tied with the univariate ensemble (adjusted $p=0.20$), TimesFM-2.5 ($p=0.17$) and Chronos-2 ($p=0.11$). It is significantly better than the rest.
\end{itemize}
In short, Chronos-2 is the best model one day ahead. At one and four weeks, a small group of foundation models and ensembles cannot be separated statistically, and every other family ranks significantly lower.

\fig{1.0}{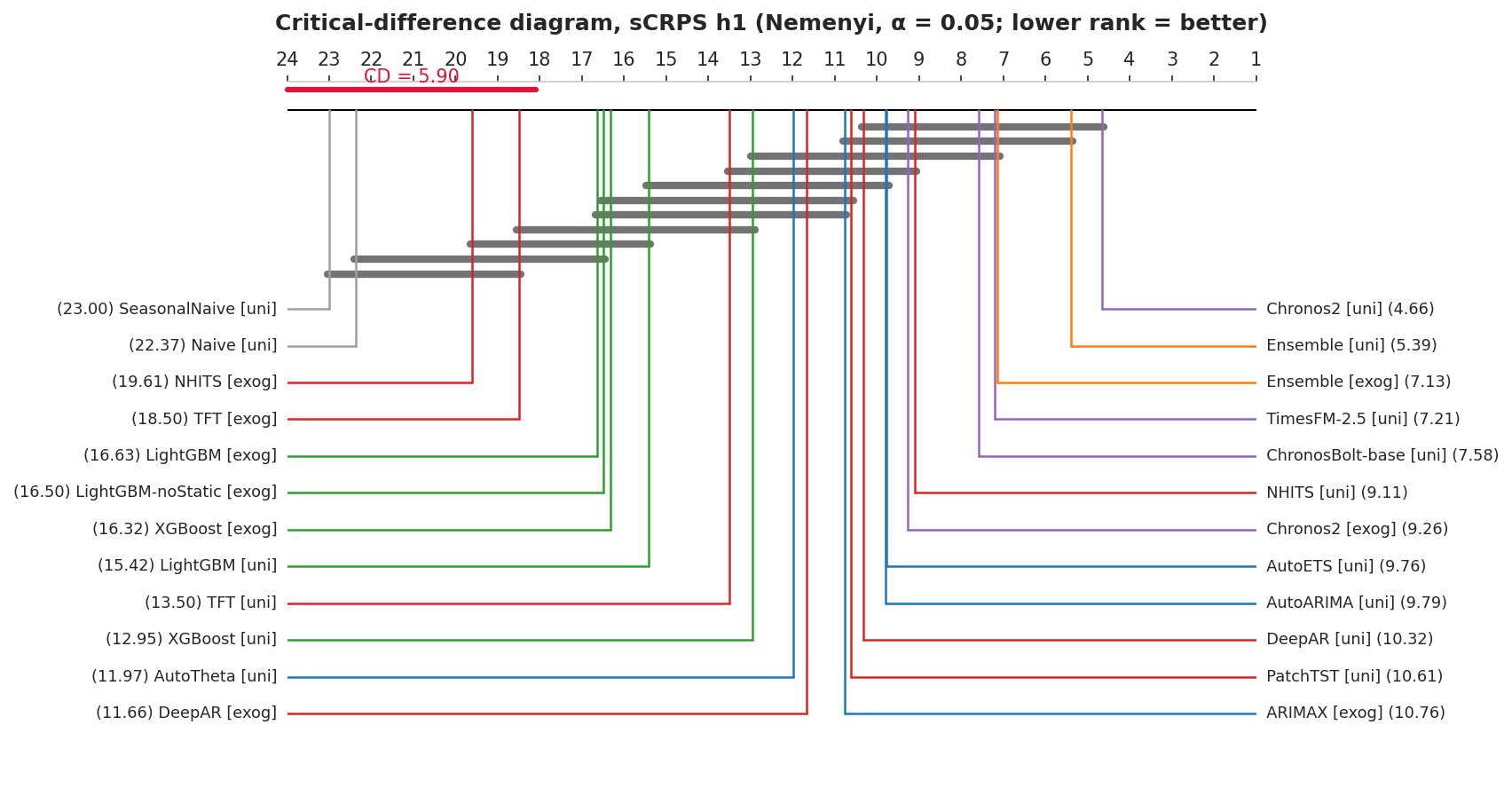}{Critical-difference diagram for sCRPS at h1 (Nemenyi, $\alpha=0.05$, CD~=~5.90). Lower mean rank is better; configurations joined by a grey bar are not significantly different.}
\fig{1.0}{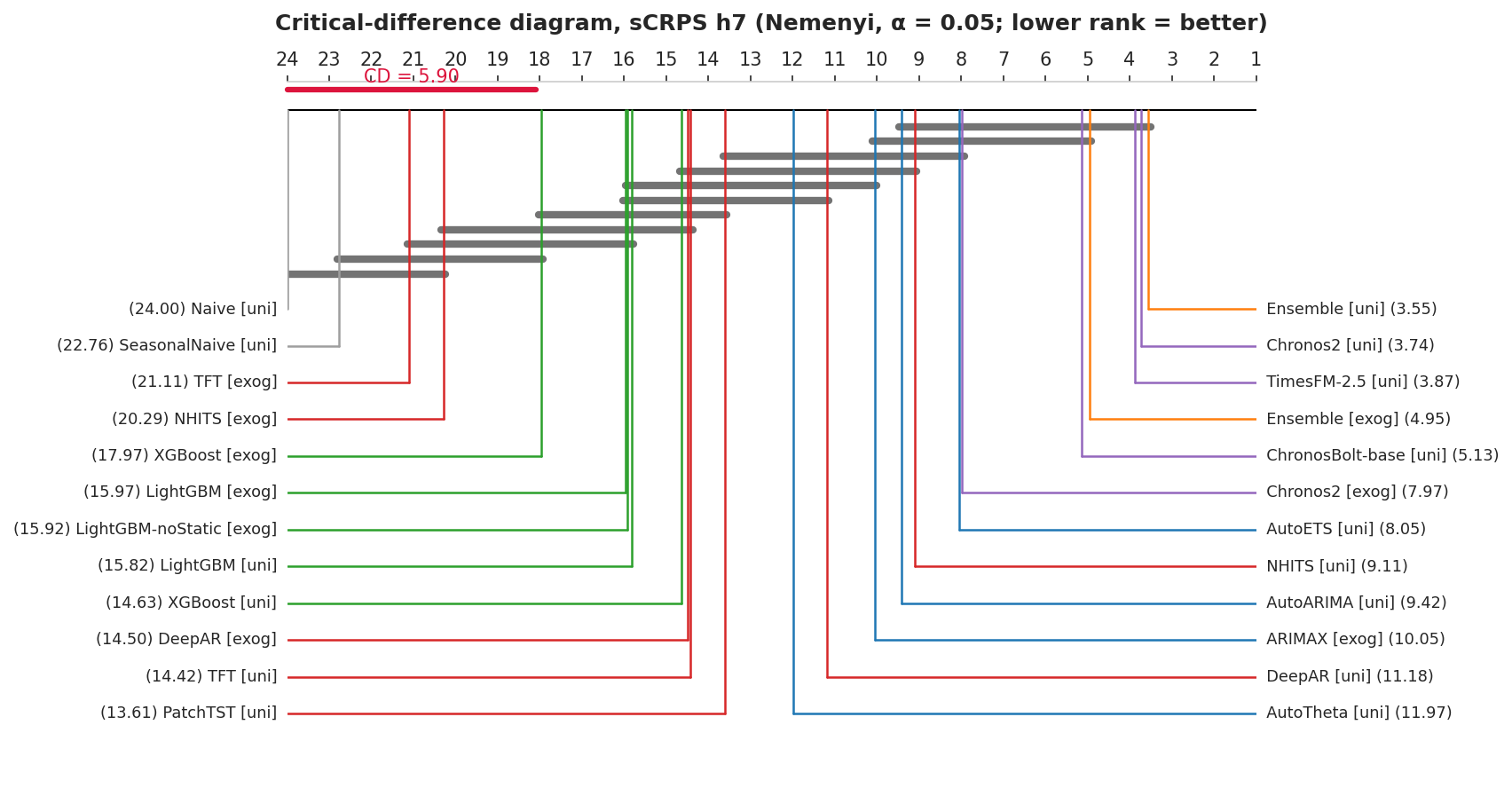}{Critical-difference diagram for sCRPS at h7.}
\fig{1.0}{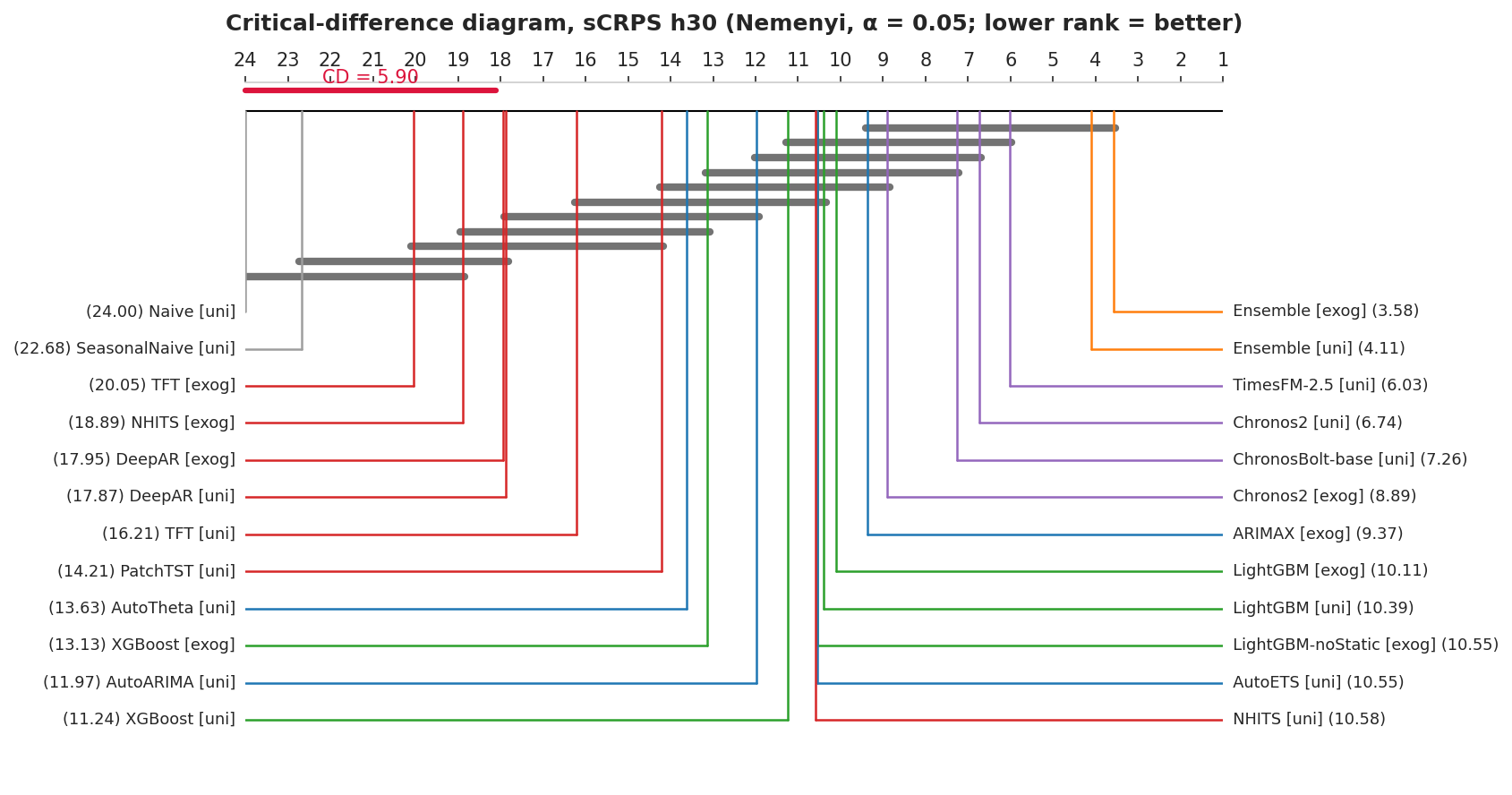}{Critical-difference diagram for sCRPS at h30.}

\subsection{Value of weather information}\label{sec:weather}
Adding the recorded weather did not improve any of the seven models that were run in both tracks. \Fig{17_weather_value_dumbbell} compares each model's univariate and weather versions at h7. The weather version is worse in every case. For the ensemble (0.674 to 0.679), LightGBM (0.698 to 0.701) and DeepAR (0.702 to 0.713) the loss is small. It is moderate for Chronos-2 (0.673 to 0.689) and XGBoost (0.688 to 0.701), and large for N-HiTS (0.691 to 0.770) and TFT (0.722 to 0.781). Measured house by house, the weather version had the lower sCRPS in 47 percent of houses for LightGBM, 26 percent for the ensemble, 16 percent for DeepAR, 11 percent for Chronos-2, 8 percent for XGBoost and in none for N-HiTS or TFT.

\fig{1.0}{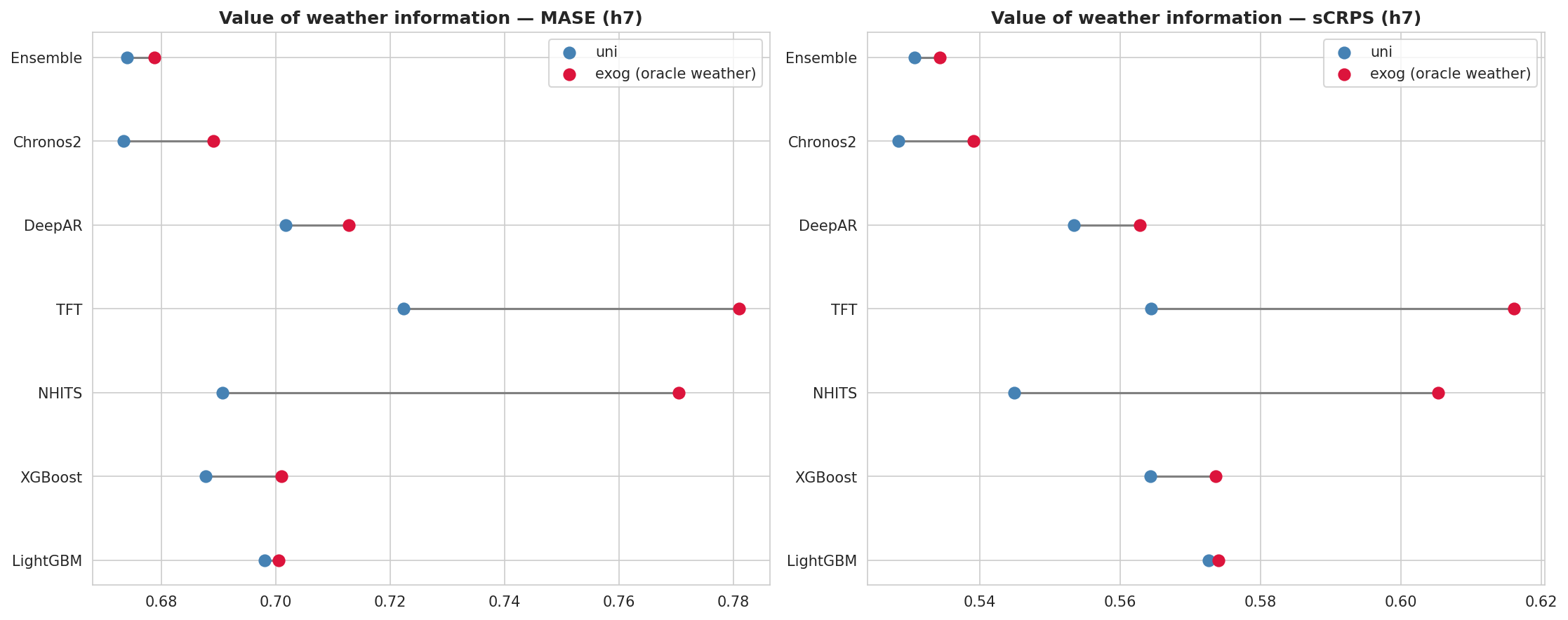}{Effect of adding the recorded weather (red) compared with the univariate version (blue) on MASE (left) and sCRPS (right) at h7. A red point to the right of its blue partner means that weather made the forecast worse. The legend label ``oracle weather'' refers to the recorded weather in the dataset, supplied for the forecast days; given Section~\ref{sec:wxqc}, it should not be read as verified observed weather.}

A test was planned of whether this result reflects the weather data rather than the physics, by scoring each pair separately on dates whose temperature passed the quality screen and on dates that did not. \Fig{30_weather_value_clean_vs_suspect} shows the outcome, but the comparison turned out to be uninformative. The flagged block of May 2024 to June 2025 covers almost the whole test period, so only one target date per house within the h7 window passed the screen (38 forecasts for each non-deep model, against 12{,}730 on suspect dates). The clean-day estimates in the figure therefore rest on a single date and should not be interpreted. The finding that does matter is the one in Section~\ref{sec:wxqc}: the weather supplied to the models for the test period is probably not the weather that occurred. If so, the weather track does not test the value of weather information at all, and the drop in accuracy of the networks that rely most heavily on these inputs is what one would expect.

\fig{0.8}{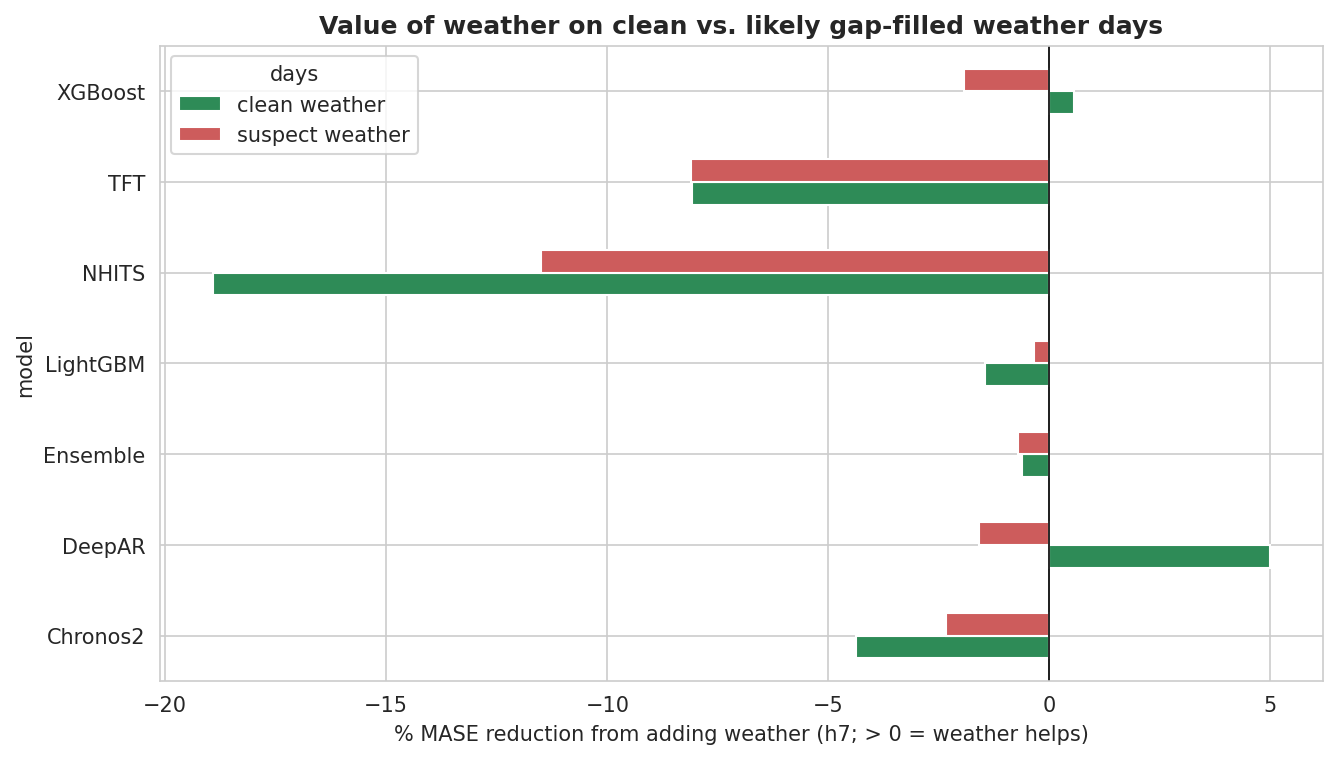}{Percentage change in MASE (h7) from adding weather, scored separately on dates whose temperature passed the quality screen (green) and on flagged dates (red). Clean dates are almost absent from the test period (one target date per house), so the green bars are not reliable estimates.}

\subsection{Value of the household survey}
Removing the survey attributes from LightGBM changed almost nothing (\Fig{18_static_ablation}). The house-by-house differences in sCRPS are tiny, at most 0.005 in any house. Their mean is $+0.0003$, $+0.0002$ and $-0.0003$ at h1, h7 and h30, and the attributes help in 47, 47 and 66 percent of houses respectively. In the gain-based importances of the trained model, all 22 survey attributes together account for 0.1 percent of the total. Two reasons are likely. First, after cleaning the survey has little left to distinguish the houses (Section~\ref{sec:eda}). Second, every house's own consumption history already encodes its level and habits far more precisely than a list of appliances can.

\fig{1.0}{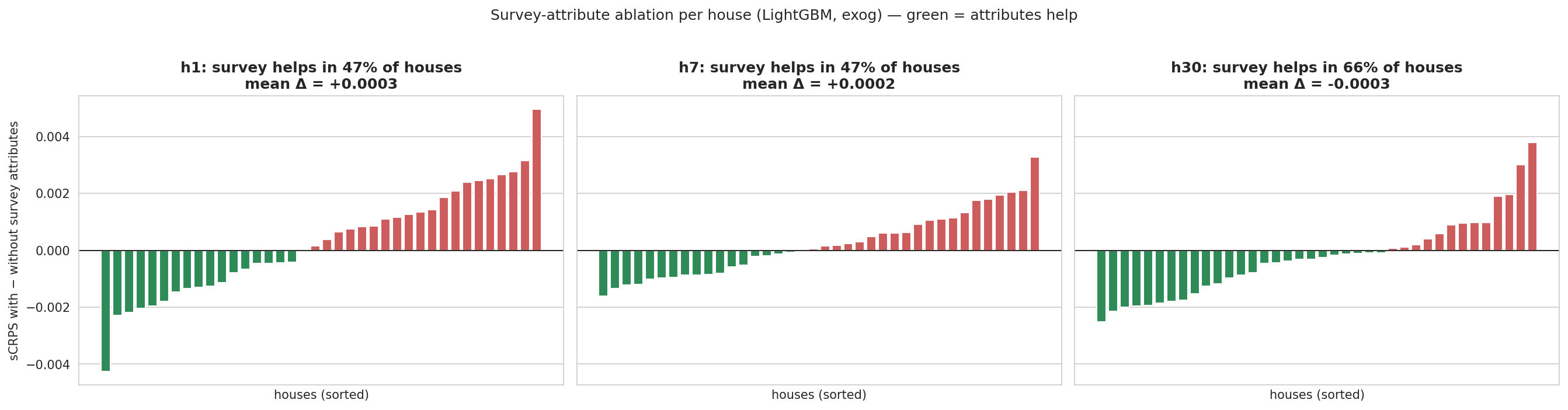}{Survey-attribute ablation for LightGBM [exog]: per-house difference in sCRPS between the model with and without survey attributes, sorted, at h1, h7 and h30. Green bars (negative values) are houses where the attributes helped.}

\subsection{What the tree models learn}
\Fig{28_gbm_feature_importance} lists the most important features of LightGBM [exog] by split gain. The 7-day rolling mean of consumption alone accounts for 75 percent of the total gain, and all rolling means together for 83 percent. Individual lags contribute 6 percent, calendar variables 4 percent and the 16 weather features together 5 percent. The most important weather feature is pressure (0.7 percent), which hardly reflects a physical driver of consumption; given Section~\ref{sec:wxqc}, it more plausibly acts as a proxy for time of year. The SHAP summary in \Fig{29_gbm_shap_summary} shows the direction of these effects. High recent consumption (red points for the 7-day mean) pushes the forecast up by as much as 1.2 standard deviations, and low recent consumption pushes it down by a similar amount. The Friday indicator has a small but consistently positive effect. Annual Fourier terms and lags at 21 and 364 days make modest contributions, and every weather variable sits close to zero. The model, in other words, has learnt to forecast mainly from the recent level of consumption, adjusted for the weekend and the season.

\fig{0.62}{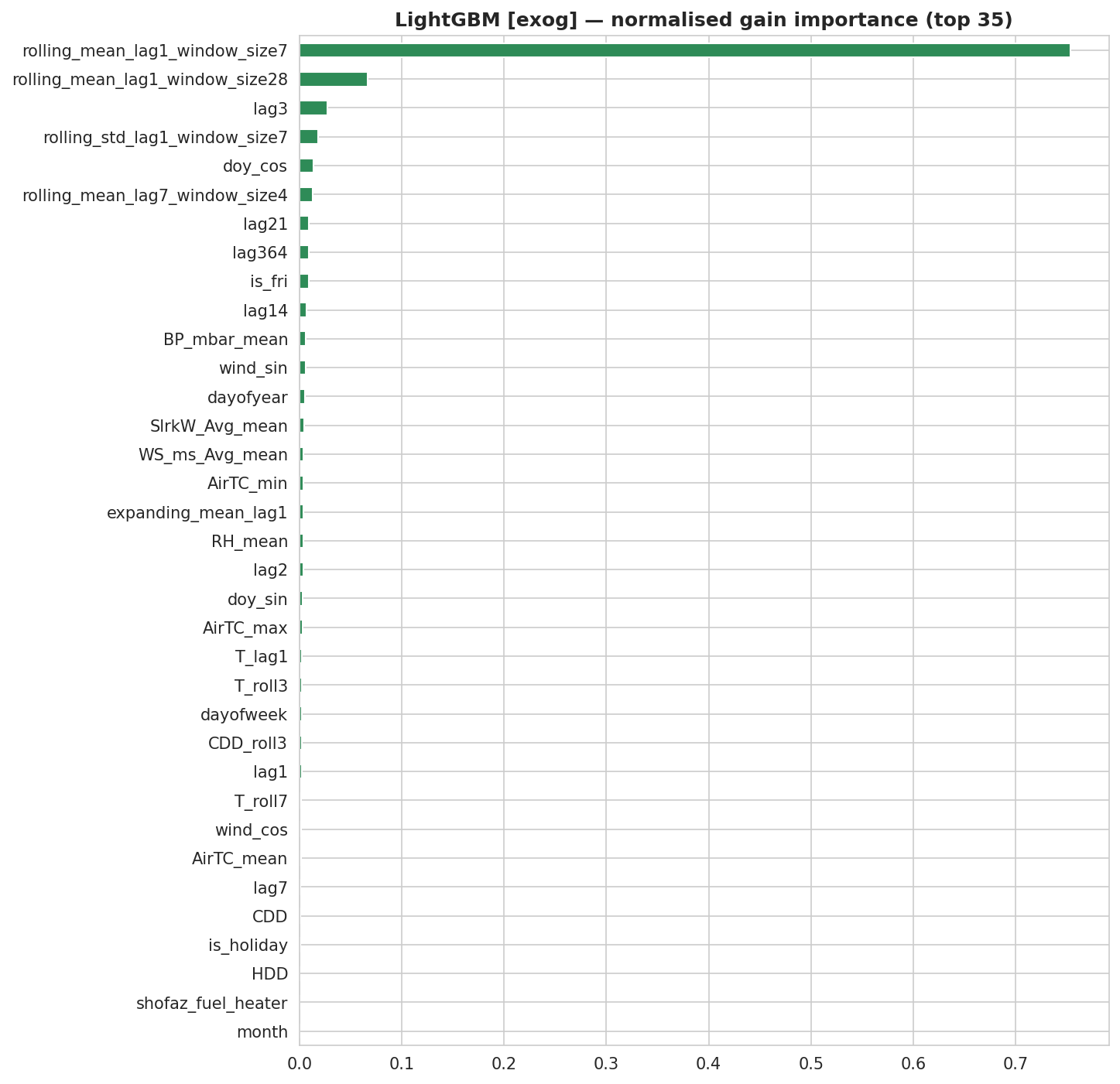}{Normalised split-gain importance of the 35 most important LightGBM [exog] features.}
\fig{0.6}{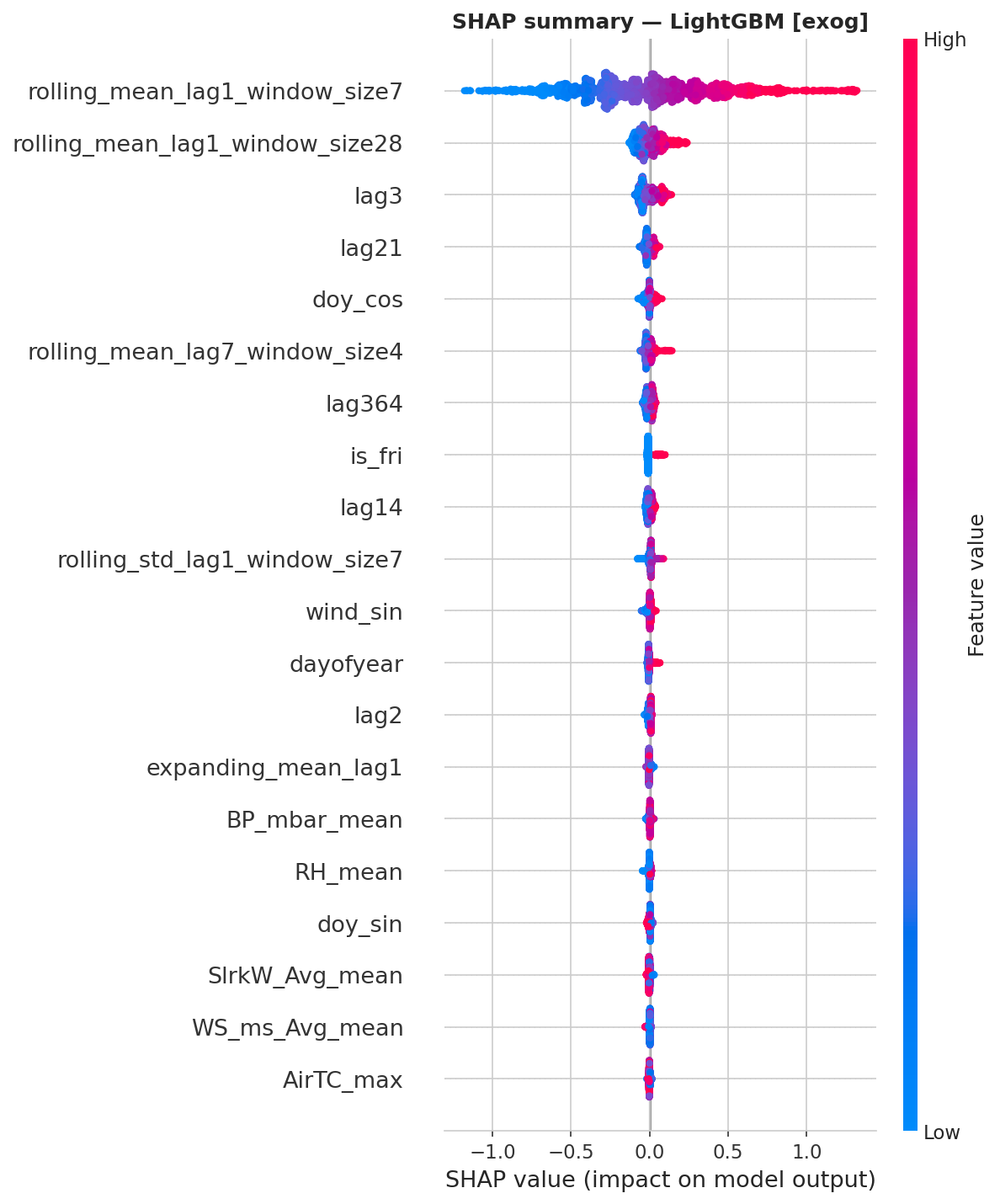}{SHAP summary for LightGBM [exog]. Each point is one forecast; horizontal position is the feature's contribution to the standardised forecast, and colour is the feature's value (red high, blue low).}

\subsection{Hyperparameter tuning}
\Fig{12b_lightgbm_tuning} traces the Optuna search. Most of the improvement came in the first few trials. The best validation MASE fell from 0.804 to 0.767 (4.6 percent) in the univariate track and from 0.823 to 0.764 (7.1 percent) in the weather track, and later trials added little. The tuned settings for both tracks are strongly regularised: 200 trees, a learning rate of about 0.014, around 30 leaves and at least 80 observations per leaf. Simple, heavily smoothed trees therefore suit this data best, which agrees with the dominance of rolling means described above.

\fig{0.72}{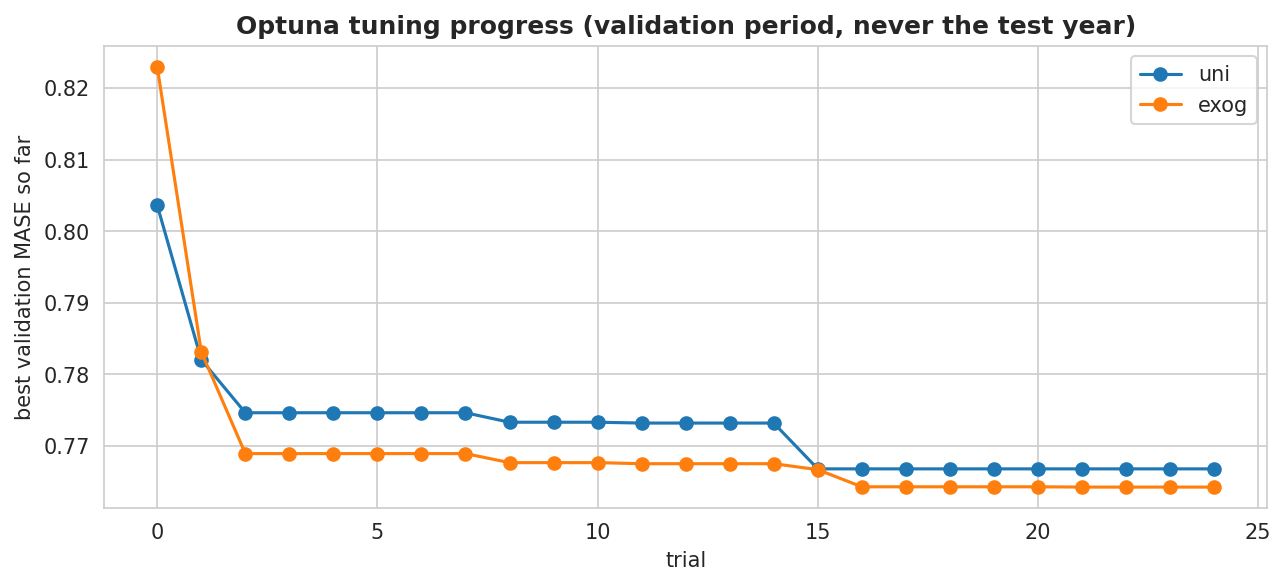}{Optuna tuning of LightGBM: best validation MASE found after each trial, for both tracks. Tuning used only the 180-day validation window.}

\subsection{Sensitivity to outage days}\label{sec:sens}
The test period contains 69 house-days below 1\,kWh. Excluding them from scoring changes the h7 MASE of every configuration by at most 0.005. The ranking is virtually unchanged: the Spearman correlation between the sCRPS rankings with and without these days is 0.997. The top three positions stay the same, with the univariate ensemble, Chronos-2 and TimesFM-2.5 at mean ranks of 3.37, 3.79 and 3.92 after exclusion. The conclusions of this study do not depend on how outage days are treated.

\subsection{Computational cost}
Accuracy and computing cost are only loosely related (\Fig{24_accuracy_vs_cost}). Over all 48 origins, Chronos-2 needed 8.5 seconds and Chronos-Bolt 11 seconds on one GPU, while ranking first and fourth by sCRPS at h7. TimesFM-2.5 took 200 seconds. Among the local models, AutoETS took 4.6 minutes, AutoARIMA 17 minutes and ARIMAX 28 minutes. The tree ensembles needed under two minutes each, plus two to three minutes of tuning. The deep networks needed between 1.3 minutes (N-HiTS [uni]) and 55 minutes (TFT [exog]) for three seeds, and TFT [exog] was the least accurate of all trained configurations. The ensembles cost the sum of their members, 21 minutes for the univariate version and 99 minutes for the weather version, most of it from ARIMAX and TFT. The foundation models therefore offer by far the best accuracy for their cost.

\fig{0.78}{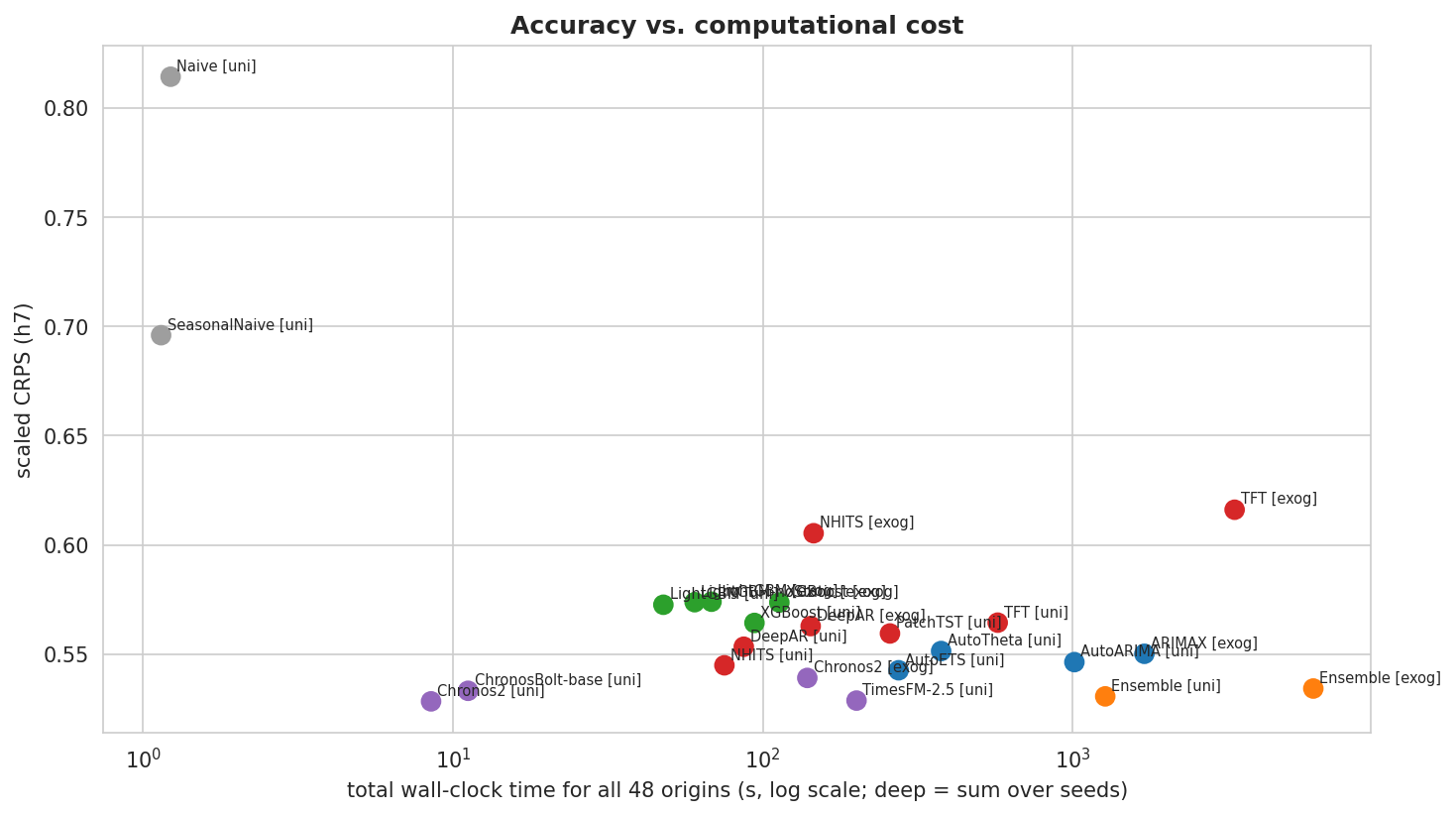}{Scaled CRPS at h7 against total wall-clock time for all 48 origins (log scale). Deep models are summed over three seeds; ensembles over their members.}

\subsection{Example forecasts}
\Fig{22_example_forecasts} shows 30-day forecasts made on 28 July 2024 for a low-, a median- and a high-consumption house. All the leading models produce nearly flat median paths at about the recent level of consumption. They make no attempt to follow day-to-day spikes, which no information available at the origin could predict. Their 80 percent intervals cover most of the observed path. The seasonal-naive forecast (grey) instead copies last week's pattern, spikes included, and its intervals widen sharply with the horizon. These examples show what forecasting skill looks like for a single home: an accurate level and an honest band, rather than an attempt to follow every peak.

\fig{1.0}{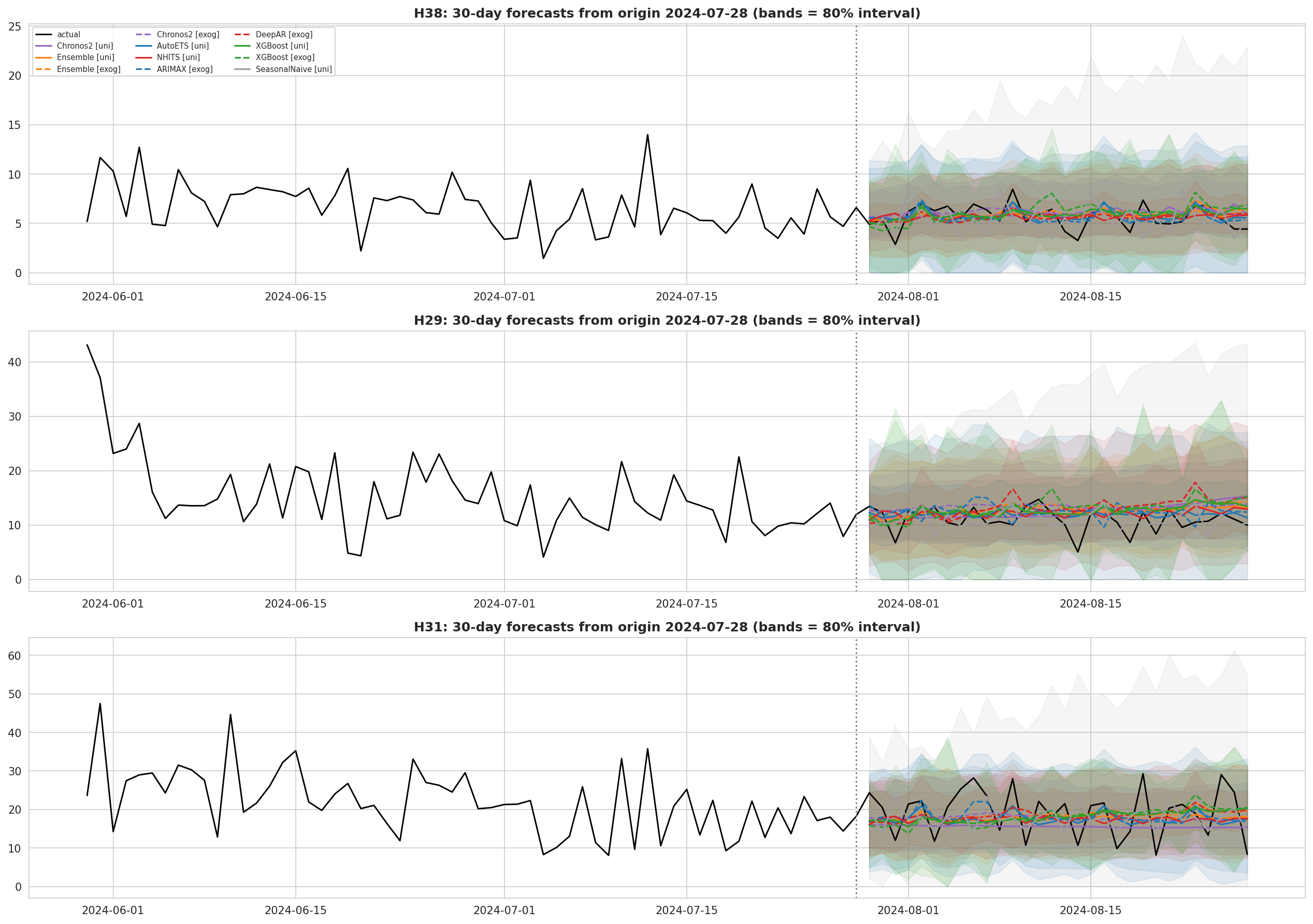}{Thirty-day forecasts from the origin of 28 July 2024 for houses H38 (low use), H29 (median) and H31 (high use). Black: observed consumption; coloured lines: median forecasts; shaded bands: 80 percent intervals.}


\section{Discussion}\label{sec:discussion}

\subsection{Why pre-trained models do so well}
Chronos-2 and TimesFM-2.5 never saw ESHED-38, and very probably no data from Iraqi homes at all. Yet Chronos-2 is the best model one day ahead, and both are statistically tied with the best ensemble at one week. The results suggest why. Daily consumption in these homes is dominated by a slowly drifting level, with a weak weekly cycle and a seasonal swing on top. The tree models' importances and SHAP values say the same: almost all of their skill comes from the recent mean. Recognising a level and placing a sensible band around it is exactly what broad pre-training on very many series teaches. The structure that specialised models are built to exploit, such as interactions with weather or household characteristics, carries little usable signal here: the survey contributes almost nothing and the weather record is unreliable. When the signal is mostly level and noise, a robust generalist is hard to beat.

The practical implication is considerable. A utility or researcher can produce forecasts of near-best accuracy, with well-calibrated intervals, for a new set of homes in seconds. No tuning, no training history and no choice of features are needed. Where computing cost is no concern, the ensemble is the safer choice beyond one day, because no other configuration is significantly better than it at one week or 30 days.

\subsection{Why deep networks trained from scratch fall short}
None of the four deep architectures beat AutoETS on average. Thirty-eight series of about 1{,}320 training days each is a small dataset for networks of this size, and the signal-to-noise ratio of single-home consumption is low. Seed-to-seed variation was small (a standard deviation of at most 0.016 in MASE at h7 and 0.021 at h30), so the gap is not an artefact of unlucky initialisation. With weather inputs, N-HiTS and TFT lost the most accuracy, and they lost it in every house. This is consistent with these networks fitting weather patterns in the training years that then failed to hold in the test period, which the repeated-sequence structure of the weather record would make very likely. These results do not show that the architectures are poor in general. They show that, on small and noisy household panels, their extra capacity is a liability unless more data or stronger constraints are available.

\subsection{Pooling across houses}
The comparison of global and local models gives a nuanced answer. At one week, pooling the 38 houses did not let the global tree ensembles or deep networks overtake AutoETS, a model fitted to each house separately: the best pooled model, N-HiTS, wins in 42 percent of houses. At 30 days the tree ensembles do reach a slightly lower relative error than AutoETS (0.808--0.809 against 0.824), but not a better sCRPS (0.776--0.780 against 0.772), and LightGBM's and AutoETS's mean ranks at that horizon are almost identical (10.4 and 10.6). Pooling across a vastly larger and unrelated corpus, as the foundation models do, did produce a consistent gain. The size and variety of the pool seems to matter more than whether the series in it are similar to the target. This agrees with the argument of Montero-Manso and Hyndman~\cite{montero2021} that the benefit of global models depends on model capacity and pool size, not on the relatedness of the series.

\subsection{What the weather results do and do not show}
The finding most likely to be misread is that recorded weather did not help. It should not be concluded from this that weather is irrelevant to household demand in Erbil. Three mechanisms may contribute. The first is redundancy: temperature changes slowly, so the past few weeks of consumption already reflect the current weather regime. The second is incomplete metering: if much summer cooling runs on generators that the meter does not record, the metered load responds to only half of the usual temperature effect. The third, and probably the decisive one, is data quality. Most of the weather record, including almost all of the test period, behaves like a reused annual sequence rather than a daily measurement. A forecaster given another year's weather for the days it must predict cannot benefit from it, and a flexible one may be misled by it. The heating-season shock of December 2024, which no model anticipated even with the recorded temperature, fits this explanation. Settling the question requires verified weather data for the same period, for instance from a national meteorological service or a reanalysis product, and this is flagged as the most important open item for anyone using ESHED-38 to study weather effects.

\subsection{Limitations}
Several limitations qualify these findings. All houses are in one city and share a single weather record, so microclimate effects cannot be studied. The survey was carried out after the metering period, so attributes may be wrong for earlier years, and it contains little variation. The meter may not capture all of a household's energy use, particularly power from generators. The test covers a single year, and the December 2024 shock shows how strongly one period can affect averages. The weather screen used here is heuristic. It identifies repeated values with high confidence, but the conclusion that whole years were reused rests on the pattern of the flags and on visual inspection, and it should be confirmed with the data owners. The weather track uses recorded weather in place of a forecast and so would overstate the value of weather even if the record were sound. Finally, with 38 houses many differences among the leading configurations are not statistically significant, and they are reported as ties.


\section{Conclusion}\label{sec:conclusion}
This paper has presented what is, to the author's knowledge, the first forecasting benchmark on ESHED-38, five years of daily electricity use for 38 homes in Erbil. Twenty-four configurations from five base model families and two cross-family ensembles were compared on identical forecasts under a rolling-origin protocol, using scale-free accuracy measures, proper probabilistic scores, calibration diagnostics and paired significance tests.

The results support five conclusions. The best methods reduce the error of a seasonal-naive rule by 20 to 25 percent and beat it in every house. Zero-shot foundation models are the most computationally efficient of the high-performing options in this benchmark: Chronos-2 is significantly the best model one day ahead and needs seconds of computation. A cross-family ensemble fixed before testing is the most robust option at longer horizons, statistically tied with the best foundation models. Deep networks trained from scratch and global tree ensembles do not consistently or significantly outperform per-house exponential smoothing on a panel of this size, and the household survey provides negligible average benefit to LightGBM. Finally, metered demand is driven by winter heating, and the weather record appears to be largely reconstructed from repeated annual sequences, which prevents any firm conclusion about the value of weather information.

Several extensions follow naturally: replacing the station record with verified weather data and with genuine weather forecasts, moving to hourly resolution, adding houses from other cities and climates, and fine-tuning foundation models on regional data. The full pipeline is released so that these extensions can be compared directly with the baselines reported here.

\section*{Data and Code Availability}
ESHED-38 is publicly available from Mendeley Data (\href{https://doi.org/10.17632/dkts3x2yh5.2}{doi:10.17632/dkts3x2yh5.2}) under a CC~BY~4.0 licence~\cite{eshed38}. The notebook that implements the complete pipeline, including data cleaning, the weather screen, all 24 configurations, evaluation and every figure, is available from the author on request.

\section*{Acknowledgements}
The author thanks Truska Jaf and Ibrahim Ismael Hamarash for making the ESHED-38 dataset openly available.


\end{document}